\documentclass[12pt]{article}
\newcommand {\al}   {\alpha}       \newcommand {\bt}  {\beta}
\newcommand {\g }   {\gamma}       \newcommand {\G }  {\Gamma}
\newcommand {\dl}   {\delta}       \newcommand {\e }  {\epsilon}
\newcommand {\z }   {\zeta}        \newcommand {\et}  {\eta}
\newcommand {\lm}   {\lambda}      \newcommand {\m }  {\mu}
\newcommand {\n }   {\nu}          \newcommand {\x }  {\xi}
\newcommand {\s }  {\sigma}
\newcommand {\ta}   {\tau}         
\newcommand {\f }   {\varphi}      \newcommand {\h }  {\chi}
         \newcommand {\om}  {\omega}

\newcommand {\pl}   {\partial}     \newcommand {\nb}  {\nabla}
\newdimen\L
\newdimen\DP
\newdimen\nnn
\newdimen\mmm
\def\cont#1#2#3{
\setbox1=\hbox{$#1$} \setbox3=\hbox{$#3$}
\setbox0=\hbox{$#1#2#3$}
\nnn=0.5\wd1 \mmm=0.5\wd3
\L=\wd0  \advance\L by -\nnn  \advance\L by -\mmm
\DP=\dp0 \advance\DP by 6pt
\hbox{%
\rlap{\box0}%
\kern\nnn\lower\DP\hbox{%
\vrule width0.4pt height3pt\vrule width\L height 0.4pt%
\vrule width0.4pt height3pt}\kern\mmm}}
\begin{document}
\title     {Canonical quantization of the string \\
            with dynamical geometry and anomaly free\\
          nontrivial string in two dimensions}
\author    {M.O.Katanaev
            \thanks{Supported by Russian fund of fundamental
		investigations RFFI--93--011--140}\\ \\
            \sl Steklov Mathematical Institute,\\
            \sl Vavilov St., 42, Moscow, 117966, Russia\\
          E-mail: katanaev@mi.ras.ru}
\date      {6 October 1993}
\maketitle
\begin{abstract}
           Hamiltonian formulation of the string with dynamical
           geometry and two-dimen\-si\-onal gravity with torsion is
           given. Canonical Hamiltonian equals to the
           linear combination of first class constraints
           satisfying closed algebra. It is
           the semidirect sum of the Virasoro algebra and
           the abelian subalgebra corresponding to the local
           Lorentz rotation. After making the canonical transformation
	     the theory is quantized. It is proved that there exists
	     Fock space representation of pure two-dimensional gravity
	     with torsion containing no central charge in the Virasoro
	     algebra. Also constructed is the new Fock representation
	     of a standard bosonic string. It is shown that two-dimensional
	     string with dynamical geometry is anomaly free and describes
	     two physical degrees of freedom.
\end{abstract}
\section{Introduction}
One of the most interesting model of contemporary physics is the
bosonic string described by an action proportional to the area
of the string world sheet (for a review see \cite{GrScWi87,BriHen88}).
At the quantum level there arises the notion of critical dimension
$D=26$ of the space-time where the theory can be consistently
quantized. This is a drawback when the theory is considered as
describing 4 dimensional extended objects. Bosonic string also
contains a tachyon in its spectrum and therefore various modifications
of the Lagrangian have been proposed last years.

The simplest geometric generalization of the bosonic string is the
string with dynamical geometry that is obtained by adding the
Lagrangian of two-dimensional gravity with torsion
\cite{KatVol86,KatVol90}. The latter equals to the sum of
curvature squared term, torsion squared term,
and cosmological constant, the geometrical quantities being
constructed from two-dimensional metric and torsion of the
string world sheet. On the other hand this model can be viewed
as two-dimensional gravity and $D$ minimally coupled massless
scalar fields corresponding to the string coordinates.

At the same time two-dimensional gravity with torsion is of interest
by itself at least for two reasons. Firstly, it provides highly
nontrivial gravity model in two dimensions with well defined action.
Therefore its elucidation will shed a new light on the whole problem
of quantum gravity. Secondly, it has direct physical application in
the theory of solids because in Euclidean space it describes the
membranes with dislocations and disclinations (see \cite{KatVol92}
and references there in).

Recently there was a progress in two-dimensional gravity
with torsion. The highly nonlinear equations of motion turn out to
possess the explicit general local solution which have been found
in the conformal gauge \cite{Katana90}, the light-cone gauge
\cite{KumSch92A}, and in the gauge where torsion components play the
role of space-time coordinates \cite{Solodu93,MiGrObTrHe93}.
Earlier there were found an instantonlike \cite{AkKiRi88} and all
static solutions \cite{KatVol90}. It is interesting that the Liouville
model (constant curvature and zero torsion) is incorporated as one of
the two sectors of two-dimensional gravity with torsion. In the
space-time equipped with general type metric and torsion satisfying
the equations of motion of two-dimensional gravity with torsion, all
extremals and geodesics have been found and analysed in the conformal
gauge \cite{Katana91}. In light-cone gauge part of them have been
analysed also in Ref.\cite{KumSch92A}. It is worth attention that
two-dimensional gravity with torsion admits not only local but also a
global analysis. Recently all global maximally continued solutions
have been found and classified up to an action of the descrete
transformation group \cite{Katana93A}. This means that at the classical
level the theory admits complete local as well as global analysis.
Among the global solutions there are solutions describing black and
white hole configurations which are similar to the Kruskal extension
of the Schwartzschild solution. The knowledge of all global solutions
of the model yields the hope to investigate in detail the relevance
of global structure of gravity to quantization.

For quantization of the model it is very important to formulate the
theory in Hamiltonian form. Two-dimensional gravity with torsion was
canonically formulated in Refs.\cite{GrKuPrSc92,Strobl93}.
(Earlier it was canonically formulated only in the conformal gauge
\cite{Katana89A}.) In Ref.\cite{GrKuPrSc92} the
lightcone variable plays the role of time. The theory is proved to
have 6 first-class costraints. The authors show that 3 of them
together with the momenta form the quadratically deformed
{\it iso(2,1)} algebra which is interpreted as the novel symmetry
of the model. In Ref.\cite{Strobl93} the theory is canonically
formulated, a timelike coordinate playing the role of time variable.
The results are in agreement with \cite{GrKuPrSc92}.
In Ref.\cite{AkDaKi92} the Hamiltonian formulation was also
considered.

Both Lagrangian and Hamiltonian formulations show that pure
two-dimensional gravity with torsion describes no dynamical
continuous degree of freedom. This statement will be surely untrue if
one adds the matter fields to the theory. This is one of the reasons
why we treat the string with dynamical geometry in the present paper.
Its canonical formulation was not considered earlier.

In the present paper the string with dynamical geometry is written
in Hamiltonian form. When one transforms a complicated nonlinear
theory with local invariance from the Lagrangian to the Hamiltonian
form the choice of independent variables become of prime importance.
In contrast to Refs.\cite{GrKuPrSc92,Strobl93} we choose to work with
the Arnowitt, Deser, Misner \cite{ArDeMi62} decomposition of the
two-dimensional metric and the corresponding parametrization of the
zweibein. This considerably simplifies calculations. We
found six first-class constraints satisfying closed algebra.
It is the direct sum of 3-dimensional abelian algebra of primary
constraints and the new 3-dimensional algebra of secondary constraints
that is the semidirect sum of the one-dimensional abelian subalgebra and
the Virasoro algebra. This is in agreement with the following general
consideration. We know that the Virasoro
algebra is connected with the conformal symmetry which is the part
of general coordinate invariance. In our case the model is also
invariant under 1-parameter local Lorentz rotation. The last symmetry
yields the nontrivial generalization of the Virasoro algebra.

The Hamiltonian form of the theory shows that the string with dynamical
geometry describes $D$ dynamical degrees of freedom. We have found
restrictions  on the coupling constants yielding positive definite
canonical Hamiltonian for all physical modes. To elucidate the quantum
constraint algebra it is very useful to make the nonlocal canonical
transformation corresponding to the decomposition of the zweibein into
longitudinal and transversal part. In fact the longitudinal component
coinsides with the conformal piece of the metric while the transversal
one coinsides with the Lorentz angle and is unphysical.

The theory is canonically quantized without any approximation.
That is for coordinates and momenta we introduce creation and
annihilation operators, construct the Fock space and prove that
the normally ordered constraints of pure two-dimensional gravity with
torsion satisfy the Virasoro algebra without central charge. We have
also found new Fock representation for an ordinary bosonic string where
the central charge differs by two from its standard value.
This means that an anomaly canarise only due to the ghost fields which
are not considered in the present paper. This is a new quantum model
describing two physical continuous degrees of freedom: the first is the
space component of the string and the second is the conformal piece
of the metric of the string world sheet.

There are other approaches to quantization of pure two-dimensional
gravity with torsion. In Refs.\cite{Strobl93,SchStr92} the theory is
canonically quantized using the symmetric representation in coordinates
and momenta of Hermitian operators. An important progress have been
done within the path integral approach. The model was shown to be
renormalisable \cite{KumSch92B} and integrable at the quantum level
\cite{HaiKum92}. The Lagrangian BRST quantization of the model was
considered in \cite{IkeIza93A}. There the action of two-dimensional
gravity with torsion was written in the first order form which
can be considered as a gauge theory of quadratically deformed
two-dimensional Poincar\'e algebra \cite{IkeIza93B}.

The paper is devided in two parts having subsections. In the
first part we discuss the classical canonical formulation of the
theory. The second part is devoted to the quantum aspects of the
theory.
\section{Hamiltonian formulation}                      \label{shamfor}
If one tries to quantize a theory nonperturbatively then the canonical
formulation seems to be the most adequate starting point. This stems
from the fact that nonlinearity of a theory does not yield any
principle difficulty when a Lagrangian theory is transformed into a
Hamiltonian form. And then it can be canonically quantized replacing
Poisson brackets by commutators. There is also not much freedom in
the introduction of creation and annihilation operators, constructing
a Fock space representation of any nonlinear quantum field theory.
There will be no approximation. The main problem arising in such
approach to gauge theories is the investigation of an algebra of
quantum operators which at the classical level satisfy some algebra
corresponding to gauge invariance. The aim here is the construction
of such representation in which the quantum algebra coincides with
the Poisson bracket algebra. In this case we say that the theory
contains no anomaly (here we neglect a possible contribution from
ghosts fields).

At the classical level a canonically formulated theory has a wide
class of equivalent realisations connected by (in general nonlinear)
canonical transformations. These realisations can yield unitary
inequivalent representations of the quantum theory. For example,
everybody knows how to
quantize a harmonic oscillator, but if one makes a nonlinear canonical
transformation then the resulting quantum theory may be disastrous
and unitary inequivalent to the standard realization. So the choice
of canonical variables becomes of prime importance for nonlinear
theories. I think the guiding principle here is the simplicity of
the resulting Hamiltonian theory were the quantization can be performed
consistently.

Situation in two-dimensional gravity with torsion is the following.
Let us find even one canonical formulation where the quantization can
be performed. This canonical formulation is given in the first part
of the paper. Here we encount four canonical transformations and
show that the resulting canonical theory for physical modes is
similar to the Liouville theory interacting with the bosonic string.
But it is more general and allows one to perform consistent
quantization.
\subsection{The Lagrangian}                             \label{slagra}
In this section we introduce notations, write down the Lagrangian and
its symmetry transformations.

Let $X^\m(\z^\al),~\m=0,1,\dots,D-1$, be coordinates of the string
moving in $D$-dimensional flat Minkowskian space-time equiped with
the metric $\et_{\m\n}=diag(+-\dots-)$. Here $\z^\al$, $\al=0,1$, are
coordinates on the string world sheet. Their range will be spesified
later, and at the moment we consider them as a local coordinate system.
To give meaning of the Hamiltonian formulation we assume that tangent vector
$\pl_0X^\m$ is everywhere timelike, $\pl_0X_\m\pl_0X^\m>0$, and
tangent vector $\pl_1X^\m$ is everywhere spacelike,
$\pl_1X_\m\pl_1X^\m<0$.

We assume that the world sheet of a string with dynamical geometry is
equiped with the Riemann--Cartan geometry, that is the metric
$g_{\al\bt}(\z)=g_{\bt\al}(\z)$ and the torsion
$T_{\al\bt}{}^\g(\z)=-T_{\bt\al}{}^\g(\z)$ are given.

The action of the string with dynamical geometry has the form
\cite{KatVol86}
\begin{equation}                                        \label{eactio}
     I=\int d^2\z e(L_X+L_G),~~~~~~~~e=\det e_\al{}^a,
\end{equation}
where $e$ defines the volume measure. Lagrangian of the model equals to
the sum of the standard Lagrangian of the bosonic string,
\begin{equation}                                        \label{elastr}
     L_X=-\frac12\rho g^{\al\bt}\pl_\al X_\m\pl_\bt X^\m,~~~~
\end{equation}
and the Lagrangian of two-dimensional gravity with torsion,
\begin {equation}                                       \label{elagra}
     L_G=\frac14 \g R_{\al\bt}{}^{ab}R^{\al\bt}{}_{ab}
         +\frac14 \bt T_{\al\bt}{}^a T^{\al\bt}{}_a - \lm,
\end {equation}
where  $ R_{\al\bt}{}^{ab} $  and  $ T_{\al\bt}{}^a $  are intrinsic
curvature and torsion of the string world sheet. Here $\rho$ is a mass
density of the string, $\g$ and $\bt$ are coupling constants and $\lm$
is a cosmological constant. Raising and lowering of Greek indices from
the beginning of the alphabet is
performed by using metric $g_{\al\bt}$ and its inverse $g^{\al\bt}$,
while Latin indices are raised and lowered by using two-dimensional
Minkowskian metric $\et_{ab}={\rm diag}(+-)$. Transformation of the
Greek indices into the Latin ones and vice versa is performed by
using zweibein field
$e_\al{}^a(\z),~g_{\al\bt}=e_\al{}^a e_\bt{}^b \et_{ab}$.

The zweibein $e_\al{}^a$ and the Lorentz connection $\om_\al{}^{ab}=
-\om_\al{}^{ba}$ are independent variables and
in this case equations of motion are of second order.
Curvature and torsion are expressed in these variables as follows
\begin {eqnarray*}
    R_{\al\bt}{}^{ab} & = & \pl_\al \om_\bt{}^{ab} - \om_\al{}^{ac}
                            \om_{\bt c}{}^b - (~\al\leftrightarrow\bt~),
\\
    T_{\al\bt}{}^a & = & \pl_\al e_\bt{}^a - \om_\al{}^{ab} e_{\bt b}
                         - (~\al\leftrightarrow\bt~).
\end {eqnarray*}

In two dimensions one can always parametrize Lorentz connection by
a pseudovector field $B_\al(\z)$
\begin{equation}                                           \label {e2}
     \om_\al{}^{ab}=B_\al\e^{ab},~~~~~~\e_{ab} = - \e_{ba},~\e_{01}=1.
\end{equation}
Then curvature takes the form
$$
	R_{\al\bt}{}^{ab}=F_{\al\bt}\e^{ab},~~~~~~
	F_{\al\bt}=\pl_\al B_\bt-\pl_\bt B_\al.
$$

In two dimensions curvature tensor is completely defined by the scalar
curvature $R$ and torsion tensor is defined by its trace $T_a$
\begin{eqnarray*}
	R_{abcd}&=&-\frac12\e_{ab}\e_{cd}R,~~~~~~~~~~R=R_{ab}{}^{ab},
\\
	T_{ab}{}^c&=&\dl_a^cT_b-\dl_b^cT_a,~~~~~~T_b=T_{ab}{}^a.
\end{eqnarray*}

By construction the action (\ref{eactio}) is invariant under general
coordinate transformations and local Lorentz rotation. To clarify
the meaning of the first class constraints in the following we need
the explicit form of the infinitesimal symmetry transformations of the
fields. Coordinate transformations with infinitesimal parameter
$\dl\z^\al=\e^\al(\z)$ look as follows
\begin{eqnarray}                                             \nonumber
	\dl X^\m &=& -\e^\bt\pl_\bt X^\m,
\\                                                           \nonumber
	\dl e_\al{}^a &=& -\pl_\al\e^\bt e_\bt{}^a-\e^\bt\pl_\bt e_\al{}^a,
\\                                                      \label{egenco}
	\dl e &=& -\pl_\al\e^\al e-\e^\al\pl_\al e
\\                                                           \nonumber
	\dl \om_\al{}^{ab} &=& -\pl_\al \e^\bt\om_\bt{}^{ab}-
	                   \e^\bt\pl_\bt\om_\al{}^{ab},
\\                                                           \nonumber
	\dl B_\al &=& -\pl_\al\e^\bt B_\bt-\e^\bt\pl_\bt B_\al.\nonumber
\end{eqnarray}
Under the Lorentz rotation by the angle $\om^{ab}=-\om^{ba}=
\om\e^{ab},~\om=-\frac12\om^{ab}\e_{ab}$, the fields transform in the
way
\begin{eqnarray}                                             \nonumber
	\dl X^\m &=& 0,
\\                                                           \nonumber
	\dl e_\al{}^a &=& \om^a{}_be_{\al}{}^{b},
\\                                                      \label{elolor}
      \dl e &=& 0,
\\                                                           \nonumber
	\dl\om_\al{}^a{}_b &=& \om^a{}_c\om_\al{}^c{}_b-
           \om_\al{}^a{}_c\om^c{}_b+\pl_\al\om^a{}_b,
\\                                                           \nonumber
	\dl B_\al &=& \pl_\al\om.
\end{eqnarray}
General coordinate transformations and local Lorentz rotation are
parametrized by three arbitrary functions and are the only local
symmetries of the string with dynamical geometry in the Lagrangian
approach \cite{Strobl93}.
\subsection{Equations of motion and the limiting cases} \label{slicas}
Before elucidating the general case corresponding to nonzero coupling
constants in the action let us say a few words about the limiting
cases. Varing the action (\ref{eactio}) with respect to $X^\m$, $B_\al$,
and $e_\al{}^a$ one obtains the following equations of motion
\begin{eqnarray}                                        \label{eqstri}
     \nb^\al\nb_\al X^\m-T^\al\nb_\al X^\m&=&0,
\\                                                      \label{eqconn}
     \g\nb_\al R+\bt T_\al&=&0,
\\                                                      \nonumber
     -\bt(\nb_\al T_\bt-\nb_\g T^\g g_{\al\bt}
     +\frac12 T_\g T^\g g_{\al\bt})-\frac14\g R^2g_{\al\bt}&&
\\                                                      \label{ezweib}
     +\rho\left(\nb_\al X_\m\nb_\bt X^\m
     -\frac12\nb_\g X_\m\nb^\g X^\m g_{\al\bt}\right)
     -\lm g_{\al\bt}&=&0,
\end{eqnarray}
where $\nb_\al$ denotes the true covariant derivative that acts with the
Lorentz connection $\om_\al{}^{ab}$ on Latin indices and the
corresponding metrical connection $\G_{\al\bt}{}^\g$ on the Greek
indices from the beginning of the alphabet. Remember that the Greek
indices from the middle of the alphabet simply enumerate $D$ string
coordinates which are scalars under changes of the coordinates on the
string world sheet.

The case $\rho=0$ corresponds to pure two-dimensional gravity with
torsion. In what follows we assume that $\rho>0$. This choice of the
sign ensure positivity of the contributions from the space string
coordinates $X^\m$, $\m\ne0$, to the canonical Hamiltonian.

For $\g=0$, equation (\ref{eqconn}) for the Lorentz connection implies
the zero torsion condition $T_{\al\bt}{}^a=0$. Then the cosmological
constant must be zero $\lm=0$ as follows from the trace of
Eq.(\ref{ezweib}). So in this case equations of motion
(\ref{eqstri})--(\ref{ezweib}) are equivalent to the equations of
motion for the ordinary bosonic string.

For $\bt=0$, trace of Eq.(\ref{ezweib}) yields the constant curvature
equation
\begin{equation}                                        \label{ecocur}
     R^2=-2F_{\al\bt}F^{\al\bt}=-4\lm/\g.
\end{equation}
Then Eq.(\ref{eqconn}) is identically satisfied, and Eqs.(\ref{eqstri}),
(\ref{ezweib}) reduce to the equations of motion for the ordinary string
for any value of the cosmological constant. Note that in this case
Eq.(\ref{ezweib}) defines metric $g_{\al\bt}$ as the one induced by the
embedding $\x^\al\rightarrow X^\m$. So the constant curvature equation
(\ref{ecocur}) must be considered with respect to the unknown Lorentz
connection. In general, torsion will differ from zero.

The general solution of pure two-dimensional gravity with torsion
contains the constant curvature and zero torsion sector
(the Liouville model) \cite{KatVol90,Katana90}
\begin{equation}                                        \label{econst}
    R^2=-4\lm/\g,~~~~T_{abc}=0.
\end{equation}
For this solution Eqs.(\ref{eqstri})--(\ref{ezweib}) reduce to the
ordinary equations for the bosonic string with one additional
restriction on the metric. On the one hand Eq.(\ref{ezweib}) implies
that the metric is induced by the embedding $\x^\al\rightarrow X^\m$.
On the other hand the corresponding scalar curvature must be constant.
That is Eq.(\ref{eqstri}) for string coordinates must be solved
for constant curvature induced metric.

In the present paper definition of the coupling constants $\bt$ and
$\g$ in Lagrangian (\ref{elagra}) differs by the sign from that in my
previous papers. This is done because in Section \ref{sthcan} we will
prove that canonical Hamiltonian for all physical modes of the theory
is positive definite if $\rho,\g,\bt>0$, and $\lm\ge 0$.
\subsection{Hamiltonian and constraints}                \label{shamco}
The Hamiltonian formulation of the Lagrangian model is the starting
point for the canonical quantization procedure. For a given Lagrangian
it is very important to choose the generalized coordinates in a
suitable way. Different choices of the coordinates are related
between themselves by the canonical transformations of the
corresponding Hamiltonian systems, and are completely equivalent
from the classical point of view. After several attempts I have found
that more convinient is the Arnowitt, Deser, Misner like
parametrization of the metric \cite{ArDeMi62}
$$
     g_{\al\bt}=g_{11}\left(
     \begin{array}{cc}
     -N_0^2+N_1^2 & N_1 \\
     N_1          & 1
     \end{array}
     \right),
$$
where $N_0\sqrt{-g_{11}}$ and $N_1$ are the lapse and shift functions.
Later we will use the conformal gauge corresponding to $N_0=1$ and
$N_1=0$. The inverse metric has the form
$$
     g^{\al\bt}=\frac1{g_{11}}\left(
     \begin{array}{cc}
     -\frac1{N_0^2}    & \frac{N_1}{N_0^2} \\
     \frac{N_1}{N_0^2} & 1-\frac{N_1^2}{N_0^2}
     \end{array}
     \right),
$$
and $\det g_{\al\bt}=-N_0^2g_{11}^2$. In accordance with our choice of
the signature we assume that $N_0>0$ and $g_{11}<0$.

The corresponding parametrization of the zweibein is
\begin{eqnarray*}
     e_0{}^a &=& (N_0e_1{}^1+N_1e_1{}^0,~N_0e_1{}^0+N_1e_1{}^1), \\
     e_1{}^a &=& (e_1{}^0,~e_1{}^1),                             \\
     \det e_\al{}^a &=& -N_0e_1{}^ae_{1a}.
\end{eqnarray*}
Therefore instead of $e_\al{}^a$ one can always choose four functions
$N_0$, $N_1$, $e_1{}^a$ as the independent variables in the action
(\ref{eactio}). This is the first canonical transformation used
in the present paper.

Let $X^\m$, $N_0$, $N_1$, $e_1{}^a$, and $B_\al$ be the generalized
coordinates in the phase space of the string with dynamical geometry.
One readily sees that no time derivatives $\pl_0N_0$, $\pl_0N_1$, and
$\pl_0B_0$ enter the Lagrangian. Therefore the corresponding conjugate
momenta $P_0$, $P_1$, and $\pi^0$ equal zero, and we have three primary
constraints
\begin{equation}                                        \label{eprico}
     P_0\approx P_1\approx\pi^0\approx0,
\end{equation}
where approximate equality means that the Poisson brackets should be
calculated before solving the constraints.
The nonzero conjugate momenta are
\begin{eqnarray}                                        \label{emomex}
      \Pi_\m &=& \frac{\dl L}{\dl(\pl_0X^\m)}=-\rho eg^{0\al}\pl_\al X_\m
\\                                                      \label{emomee}
      \pi^1{}_a &=& \frac{\dl L}{\dl(\pl_0e_1{}^a)}=
                    \bt eT^{01}{}_a=\bt\e_{ab}T^b,
\\                              			        \label{emomeb}
      \pi^1 &=& \frac{\dl L}{\dl(\pl_0B_1)}=-2\g eF^{01}=-\g R.
\end{eqnarray}

We see that the theory contains 3 primary constraints. This is the
minimal number
because each local symmetry of the Lagrangian must be accompanied by
one primary constraint \cite{HeTeZa90}.

After tedious but straightforward calculations one obtains the
canonical Hamiltonian
\begin{eqnarray}                                           \nonumber
	H &=& \int d\s\left(\Pi_\m\pl_0X^\m+\pi^1{}_a\pl_0e_1{}^a+
	  \pi^1\pl_0B_1-L\right)
\\                                                    \label{ehamil}
        &=& \int d\s\left(N_0{\cal H}_0+N_1{\cal H}_1+B_0\h\right),
\end{eqnarray}
where
\begin{eqnarray}                                        \label{econs1}
    {\cal H}_0&=&-\frac1{2\rho}\Pi_\m\Pi^\m-\frac12\rho\pl_1X_\m\pl_1X^\m
    +g_{11}E-(\pl_1\pi^1{}_a+B_1\pi^1{}_b\e^b{}_a)\e^{ac}e_{1c},
\\                                                      \label{econs2}
    {\cal H}_1&=&\Pi_\m\pl_1X^\m
    -(\pl_1\pi^1{}_a+B_1\pi^1{}_b\e^b{}_a)e_1{}^a,
\\                                                      \label{econs3}
    \h&=&-\pl_1\pi^1+\pi^1{}_a\e^{ab}e_{1b},
\end{eqnarray}
and we have used the abbreviation
\begin{equation}                                        \label{edefen}
    E=\frac1{2\bt}\pi^1{}_a\pi^{1a}-\frac1{4\g}\pi^1\pi^1-\lm.
\end{equation}

Equal time Poisson brakets between conjugate variables in points $\s$ and
$\s'$ have the usual form
\begin{eqnarray*}
	\{X^\m,\Pi'_\n\} &=& \dl_\n^\m\dl(\s-\s'),
\\
	\{N_0,P_0'\}     &=& \dl(\s-\s'),
\\
	\{N_1,P_1'\}     &=& \dl(\s-\s'),
\\
	\{e_1{}^a,\pi'^1{}_b\} &=& \dl_b^a\dl(\s-\s'),
\\
	\{B_\al,\pi'^\bt\} &=& \dl_\al^\bt\dl(\s-\s').
\end{eqnarray*}
Here primes denote the space point at which the variables are considered.

The functions ${\cal H}_0$, ${\cal H}_1$, and $\h$ do not depend on
$N_0$, $N_1$, and $B_0$. Therefore the secondary constraints are easily
obtained considering equations of motion of primary constraints
\begin{eqnarray}                                        \nonumber
    \pl_0P_0&=&\{P_0,H\}=-{\cal H}_0\approx0,
\\                                                      \label{esecon}
    \pl_0P_1&=&\{P_1,H\}=-{\cal H}_1\approx0,
\\                                                      \nonumber
    \pl_0\pi^0&=&\{\pi^0,H\}=-\h\approx0.
\end{eqnarray}

So the theory contains 3 secondary constraints, and the Hamiltonian
(\ref{ehamil}) equals to their linear combination. Therefore evolution
of the secondary constraints is entirely defined by the Poisson bracket
algebra of the constraints. Straightforward calculations show that the
secondary constraints satisfy the following algebra
\begin{eqnarray}                                             \nonumber
	\{{\cal H}_0,{\cal H}'_0\} &=& ({\cal H}_1+{\cal H}'_1)\dl'(\s'-\s),
\\                                                           \nonumber
	\{{\cal H}_0,{\cal H}'_1\} &=& ({\cal H}_0+{\cal H}'_0)\dl'(\s'-\s)
	+\frac14g_{11}\pi^1\h\dl(\s'-\s),
\\                                                         \label{epoalg}
	\{{\cal H}_1,{\cal H}'_1\} &=& ({\cal H}_1+{\cal H}'_1)\dl'(\s'-\s),
\\                                                           \nonumber
	\{\h,{\cal H}'_0\} &=& \{\h,{\cal H}'_1\}=\{\h,\h'\}=0,
\end{eqnarray}
where primes on constraints denote their argument $\s'$, while $\dl'$
denotes the derivative of the $\dl$-function,
$$
     \dl'(\s'-\s)=\pl_\s\dl(\s'-\s).
$$

It is clear also that the primary constraints commute between themselves
and with the secondary constraints.

So the primary and secondary constraints are of the first class in
Dirac's terminology. The Hamiltonial system is in involution, and
there is no need to compute tertiary constraints.

The obtained Hamiltonian equals to the linear combination of the
first class constraints. This is a common feature of the theories
invariant under general coordinate transformation. In contrast to
general relativity all constraints are polinomial in generalized
coordinates and momenta. This is a particular feature of two dimensions.

Let us briefly discuss the geometrical meaning of the constraints.
The integral
$$
	Q_L=-\int d\s\om\h,
$$
where $\om$ is a parameter of local Lorentz rotation, generates correct
Lorentz rotation (\ref{elolor}) for the fields $e_1{}^a$, $B_1$,
and $X^\m$. For example,
$$
	\dl B_1=\{B_1,Q_L\}=\pl_1\om.
$$
In the same way one may check that the integral
$$
	Q_1=-\int d\s\e^1{\cal H}_1
$$
generates coordinate transformations (\ref{egenco}) for the same
fields with the parameter $\dl\z^\al$$=\e^1\dl_1^\al$. The integral
$$
	Q_0=-\int d\s\e^0{\cal H}_0
$$
is necessary to generate correct transformation of the time coordinate
$\dl\z^\al$$=\e^0\dl_0^\al$. In the latter case one must exclude
time derivatives from (\ref{egenco}) using the equations of motion,
which in their turn are defined by the Hamiltonian. We do not discuss
this subtle point here and only note that in the conformal gauge
$N_0=1$, $N_1=B_1=0$ the constraint ${\cal H}_0$ represents the energy
density.
\subsection{The algebra of the constraints}             \label{salcon}
The Poisson bracket algebra of the constraints (\ref{epoalg}) is not
a true algebra because there are structure functions instead of
structure constants. Let us pose the question whether there exists
another equivalent set of constraints forming a true algebra?
This question was solved in \cite{GrKuPrSc92} by considering more
general algebra including the momenta besides the constraints. The
algebra was shown to be that of quadratically deformed
$iso(2,1)$-algebra. However there does exist the equivalent set of
constraints by themselves satisfying the closed algebra. We show
that this algebra is a semidirect sum of the Virasoro algebra
and the one dimensional abelian algebra corresponding to the
Lorentz rotation.

Let us define the new set of constraints
\begin{eqnarray}                                        \nonumber
    \widetilde{\cal H}_0 &=& {\cal H}_0+B_1\h,
\\                                                      \label{enecon}
    \widetilde{\cal H}_1 &=& {\cal H}_1+B_1\h,
\end{eqnarray}
the constraint $\h$ being unchanged. Then straightforward calculations
show that the new algebra closes with the structure constants instead
of structure functions
\begin{eqnarray}                                        \nonumber
    \{\widetilde{\cal H}_0,\widetilde{\cal H}'_0\} &=&
    (\widetilde{\cal H}_1+\widetilde{\cal H}'_1)\dl'(\s'-\s),
\\                                                      \nonumber
    \{\widetilde{\cal H}_0,\widetilde{\cal H}'_1\} &=&
    (\widetilde{\cal H}_0+\widetilde{\cal H}'_0)\dl'(\s'-\s),
\\                                                      \nonumber
    \{\widetilde{\cal H}_1,\widetilde{\cal H}'_1\} &=&
    (\widetilde{\cal H}_1+\widetilde{\cal H}'_1)\dl'(\s'-\s),
\\                                                      \label{enealg}
    \{\h,\widetilde{\cal H}'_0\} &=& \h'\dl'(\s'-\s),
\\                                                      \nonumber
    \{\h,\widetilde{\cal H}'_1\} &=& \h'\dl'(\s'-\s),
\\                                                      \nonumber
    \{\h,\h'\} &=& 0.
\end{eqnarray}
The first three Poisson brackets show that $\widetilde{\cal H}_0$ and
$\widetilde{\cal H}_1$ form the Virasoro or the conformal algebra.
The abelian constraint $\h$ form an invariant subalgebra under the
action of the Virasoro generators. So the new algebra is a semidirect
sum of the Virasoro algebra and the one-dimensional abelian algebra.
This algebra corresponds to the semidirect product of the conformal
symmetry transformations on the Lorentz rotation.

Let us rewrite the algebra (\ref{enealg}) in other forms to compare
it with the Fourier transformed Virasoro algebra. Introducing
generators
$$
    {\cal H}^\pm=\frac12(\widetilde{\cal H}_0\pm\widetilde{\cal H}_1),
$$
the algebra takes the form
\begin{eqnarray}                                        \nonumber
    \{{\cal H}^+,{\cal H}'^+\} &=&
    ({\cal H}^++{\cal H}'^+)\dl'(\s'-\s),
\\                                                      \nonumber
    \{{\cal H}^+,{\cal H}'^-\} &=& 0,
\\                                                      \label{eplalg}
    \{{\cal H}^-,{\cal H}'^-\} &=&
    -({\cal H}^-+{\cal H}'^-)\dl'(\s'-\s),
\\                                                      \nonumber
    \{\h,{\cal H}'^+\} &=& \h'\dl'(\s'-\s),
\\                                                      \nonumber
    \{\h,{\cal H}'^-\} &=& \{\h,\h'\} = 0.
\end{eqnarray}
We see that the algebra generated by ${\cal H}^-$ completely
decouples, and we are left with the semidirect sum of the
algebras generated by ${\cal H}^+$ and $\h$.

The asymmetry between ${\cal H}^+$ and ${\cal H}^-$ is insignificant.
If one redifines the generator $\widetilde{\cal H}_0$ in (\ref{enecon})
using the expression $\widetilde{\cal H}_0={\cal H}_0-B_1\h$ and leaves
the definition of $\widetilde{\cal H}_1$ and ${\cal H}^\pm$ unchanged,
then decoupled is the algebra generated by ${\cal H}^+$. This shows that
the symmetry between ${\cal H}^+$ and ${\cal H}^-$ in
(\ref{eplalg}) was broken by the definition (\ref{enecon}).

Let us make the Fourier transform
\begin{eqnarray*}
	L_m^\pm &=& \int_{-\pi}^\pi d\s{\cal H}^\pm e^{-\imath m\s},
\\
	M_m &=& \int_{-\pi}^\pi d\s\h e^{-\imath m\s}.
\end{eqnarray*}
Then the algebra looks as follows
\begin{eqnarray}                                             \nonumber
	\{L_m^+,L_n^+\} &=& \imath(m-n)L_{m+n}^+,
\\                                                           \nonumber
	\{L_m^-,L_n^-\} &=& -\imath(m-n)L_{m+n}^-,
\\                                                      \label{evirco}
	\{L_m^+,M_n\} &=& \imath mM_{m+n},
\\                                                           \nonumber
	\{L_m^+,L_n^-\} &=& \{L_m^-,M_n\}=\{M_m,M_n\}=0.
\end{eqnarray}

Thus we have analysed the algebraic structure of the constraints of
the string with dynamical geometry. The theory contains 6 first class
constraints. Three primary first class constraints are abelian and
completely decouple from the algebra, while 3 secondary first class
constraints form the nontrivial generalization of the Virasoro algebra.
\subsection{Extra constraint in two-dimensional gravity with torsion}
                                                        \label{sexcon}
In the case of pure two-dimensional gravity with torsion there is one
extra first class constraint. In this section we assume that the string
coordinates are absent.

Direct integration of the equations of motion of two-dimensional gravity
with torsion \cite{Katana90,KumSch92A} shows that there is the invariant
relation between scalar curvature and torsion
\begin{equation}                                        \label{einvre}
    \left(\frac{\g^2}{\bt^2}R^2+2\frac\g\bt R+2
    +\frac{4\lm\g}{\bt^2}+\frac\g\bt T_{abc}T^{abc}\right)e^{-\g R/\bt}=A,
\end{equation}
where $A$ is an arbitrary constant (integral of motion) defined by the
initial data. This relation holds for any solution of the equations of
motion. Comparing (\ref{einvre}) with the definition of the momenta
(\ref{emomee}) and (\ref{emomeb})
we see that the integral of motion can be expressed intirely in terms
of the canonical momenta
\begin{equation}                                        \label{extcon}
    A=-\frac{4\g}{\bt^2}\left(E+\frac\bt{2\g}\pi^1-\frac{\bt^2}{2\g}\right)
    e^{\pi^1/\bt}
\end{equation}
where $E$ is defined by (\ref{edefen}).
Because A is the integral of motion and the Hamiltonian equals to the
linear combination of the first class constraints, the expression
(\ref{extcon}) must be a first class constraint. Indeed, straightforward
calculations yield the following Poisson brackets
\begin{eqnarray*}
	\{A,{\cal H}'^+\} &=& \frac{4\g}{\bt^3g_{11}}
	(\pi^{1a}e_{1a}-\pi^{1a}\e_{ab}e_1{}^b)e^{\pi^1/\bt}{\cal H}^+
	\dl(\s'-\s)
\\
	&+& \frac{4\g}{\bt^3g_{11}}\left[g_{11}E-B_1
	(\pi^{1a}e_{1a}-\pi^{1a}\e_{ab}e_1{}^b)\right]
	e^{\pi^1/\bt}\h\dl(\s'-\s)
\\
	\{A,{\cal H}'^-\} &=& \frac{4\g}{\bt^3g_{11}}
	(\pi^{1a}e_{1a}+\pi^{1a}\e_{ab}e_1{}^b)e^{\pi^1/\bt}
	{\cal H}^-\dl(\s'-\s)
\\
	\{A,\h'\} &=& \{A,A'\}=0.
\end{eqnarray*}

At first sight $A$ is an extra independent first class constraint of
two-dimensional gravity with torsion. But this is not so. One may
check that its space derivative is the linear combination of the
secondary constraints for $X^\m=0$
\begin{equation}                                        \label{extder}
     \pl_1A=\frac{4\g}{\bt^3}\left(
     \frac{e_1{}^a\e_{ab}\pi^{1b}}{g_{11}}{\cal H}_0
     +\frac{e_{1a}\pi^{1a}}{g_{11}}{\cal H}_1+E\h\right)e^{\pi^1/\bt}
\end{equation}
So it does not generate new gauge symmetry \cite{Strobl93}. In fact,
explicit integrability of the right hand side of (\ref{extder}) is
connected to the integrability of pure two dimensional gravity with
torsion.
\subsection{The second canonical transformation}        \label{scatra}
Having in hand the Hamiltonian theory one can try to quantize it.
By this we mean the explicit construction of the Fock space where
all the operators will work. In Ref.\cite{SchStr92} two-dimensional
gravity with torsion is quantized using other representation of the
operators with a symmetric ordering of the coordinates and momenta
without introduction of creation and annihilation operators.
Hopefully, different schemes of quantization will add new features
to the theory.

To construct the Fock space one needs to introduce creation and
annihilation operators. This can be done in the full nonlinear
theory through the coordinates and momenta in the usual way without
referring to the equations of motion. Then
all the operators including the constraints must be supplemented by
the normal ordering prescription. At this point one faces the
problem of quantum anomalies because the algebra of the constraints
must be closed also at the quantum level. Remember that in the theory
of ordinary bosonic theory the anomaly exists. It is the central
charge of the Virasoro algebra which is compensated by the ghost
contribution in $D=26$.

In the case of the string with dynamical geometry the constraints are
polinomials of the fourth order in creation and annihilation operators,
and the calculation of the anomalies in the algebra can be, in principle,
performed. But I failed to do this. Either the calculations become
unmanagable or the central charge diverge. This problem forces me to
make three more canonical transformations before the introduction of the
creation and annihilation operators. The first was already done when we
chose the canonical coordinates in the Lagrangian and here we consider
the second most important canonical transformation.

To simplify the form of the constraints let us make the second nonlocal
canonical transformation
$$
     e_1{}^a,\pi^1{}_a,B_1,\pi^1\rightarrow Q,Q_\perp,P,P_\perp,Q_B,P_B
$$
corresponding to the separation of the longitudinal and transversal
part of the zweibein. The string coordinates and momenta remain
unchanged. May be this very peculiar canonical transformation is the
most significant technical achievement of the present paper. We choose
the generating function depending on old coordinates and new momenta
\begin{equation}                                        \label{ecatrf}
	F_2=P\frac{e_1{}^a\pl_1e_{1a}}{g_{11}}+P_\perp\frac{e_1{}^a\e_{ab}
	\pl_1e_1{}^b}{g_{11}}-P_\perp B_1+P_BB_1.
\end{equation}
Then one can find the expressions for old momenta and new coordinates
\begin{eqnarray}                                          \nonumber
     \pi^1{}_a &=& \frac{\dl F_2}{\dl e_1{}^a}=
     -\pl_1P\frac{e_{1a}}{g_{11}}-\pl_1P_\perp\frac{e_1{}^b\e_{ba}}{g_{11}},
\\                                                        \nonumber
    \pi^1 &=& \frac{\dl F_2}{\dl B_1}=-P_\perp+P_B,
\\                                                      \label{enewqp}
    Q &=& \frac{\dl F_2}{\dl P}=\frac{e_1{}^a\pl_1e_{1a}}{g_{11}},
\\                                                        \nonumber
    Q_\perp &=& \frac{\dl F_2}{\dl P_\perp}
    =\frac{e_1{}^a\e_{ab}\pl_1e_1{}^b}{g_{11}}-B_1,
\\                                                        \nonumber
    Q_B &=& \frac{\dl F_2}{\dl P_B}=B_1.
\end{eqnarray}
The first of the above equations show that $\pl_1P$ and $\pl_1P_\perp$
represent the longitudinal and transversal part of $\pi^1{}_a$
correspondingly. The appearence of the partial derivative $\pl_1$ in
(\ref{enewqp}) means that the canonical transformation is nonlocal.
It will be shown however that the nonlocality is inessential when the
theory is considered on a cylinder corresponding to a closed string.

Calculating the right hand side of (\ref{enewqp}) we use the identity
$$
    \frac\dl{\dl e_1{}^a}\int d\s\frac{e_1{}^a\e_{ab}\pl_1e_1{}^b}{g_{11}}
    =0.
$$
This piculiar identity is proved by direct calculation and is nontrivial
because the integrand is not equal to a total divergence.

The second used identity
$$
    \frac\dl{\dl e_1{}^a}\int d\s\frac{e_1{}^a\pl_1e_{1a}}{g_{11}}=0
$$
is trivial because the integrand is a total divergence
$$
    \frac{e_1{}^a\pl_1e_{1a}}{g_{11}}=\frac12\pl_1\ln(-g_{11}),
    ~~~~~~~~g_{11}<0.
$$

Using Eqs.(\ref{enewqp}) the constraints can be written in the form
\begin{eqnarray}                                           \nonumber
     {\cal H}_0 &=&-\frac1{2\rho}\Pi_\m\Pi^\m-\frac12\rho\pl_1X_\m\pl_1X^\m
\\                                                         \nonumber
     &+& \frac1{2\bt}(\pl_1P)^2-\frac1{2\bt}(\pl_1P_\perp)^2
     -\pl_1PQ_\perp-\pl_1P_\perp Q
     +\pl_1^2P_\perp-\frac1{4\g}g_{11}(P_\perp-P_B)^2-\lm g_{11},
\\                                                      \label{ecosec}
     {\cal H}_1 &=&
     \Pi_\m\pl_1X^\m-\pl_1PQ-\pl_1P_\perp Q_\perp+\pl_1^2P,
\\                                                         \nonumber
     \h &=& -\pl_1P_B.
\end{eqnarray}
The constraint ${\cal H}_0$ is quadratic in the fields except the
linear term and two last terms containing nonpolinomial space
component of the metric
\begin{equation}                                        \label{emetsp}
	g_{11}=-\exp\left(2\int_{\s_0}^\s d\s'Q(\ta,\s')\right).
\end{equation}
This is the consequence of the nonlocality of the canonical
transformation. The constant of integration $\s_0$ is just the point
where $g_{11}$$=-1$. So at this point the unit of space length is chosen
to be unity at every moment of time. The second constraint
${\cal H}_1$ contains quadratic and one linear terms. The last
constraint $\h$ is linear.

To clarify the geometric meaning of the second canonical transformation
let us paramet\-ri\-ze the zweibein components $e_1{}^a$ by the Lorentz
angle $\om$ and the conformal factor $e^\f>0$, where $\f(\x)$ is a
"scalar" field. Indeed, $e_1{}^a$ is a Lorentz vector
and can be always parametrized as follows
\begin{equation}                                        \label{ezwpar}
     e_1{}^a=e^\f(n^a\cosh \om+\e^{ab}n_b\sinh\om),~~~~~~~~
     g_{11}=-e^{2\f}.
\end{equation}
where $n^a$ is a fixed spacelike vector of unit length $n^an_a=-1$.
Then Eqs.(\ref{enewqp}) imply
\begin{equation}                                        \label{esecpa}
     Q=\pl_1\f,~~~~~~~~Q_\perp=\pl_1\om.
\end{equation}
So longitudinal and transversal components $Q$, $Q_\perp$ are
parametrized by the comformal factor and the Lorentz angle.
The important point is that the initial theory is invariant under
local Lorentz rotation. This sugests that the modes corresponding to
$Q_\perp$, $P_\perp$ are unphysical and can be always gauged away.

The second canonical transformation is not a one-to-one transformation.
The first equation (\ref{enewqp}) shows that $P$ and $P_\perp$ are
defined through the original coordinates and momenta up to an addition
of arbitrary constants. So dealing with new canonical variables one can
add arbitrary constants to $P$ and $P_\perp$ and the corresponding
constant to $P_B$ without changing the initial theory. This freedom
will be used in the following section. At the same time the new set of
canonical variables also does not uniquely define the original
coordinates. The conformal factor $\f$ and the Lorentz angle $\om$ are
defined by (\ref{esecpa}) up to an arbitrary constants too.
\subsection{Fixing of the gauge}                        \label{sfigau}
According to a general prescription one can introduce creation and
annihilation operators for all fields describing both physical and
unphysical degrees of freedom. Then one must check whether the algebra
of the constraints is satisfied at the quantum level too.

String with dynamical geometry is described by $D+6$ pairs of
canonically conjugate variables. Six first class constraints
together with six gauge conditions can be used to eliminate
6 unphysical degrees of freedom. So the string with dynamical geometry
contains $D$ physical degrees of freedom.

To quantize the model we use the gauge freedom only partially. Namely,
we eliminate 4 unphysical degrees of freedom explicitly and leave 2
unphysical degrees of freedom and 2 first class constraints
satisfying the pure Virasoro algebra.
This has been done to simplify the quantization procedure and to be
as close to the ordinary bosonic string theory as possible.

The extended Hamiltonian of the string with dynamical geometry equals
to the linear combination of all constraints
\begin{equation}                                        \label{extha}
    H_E=\int d\s(v_0P_0+v_1P_1+v\pi^0+u_0{\cal H}_0
    +u_1{\cal H}_1+u\h),
\end{equation}
where $v_0,v_1,v,u_0,u_1,u$ are Lagrange multiplyers. According to a
general prescription one can choose one gauge condition for every
first class constraint. Every gauge condition together with the
corresponding first class constraint must form a pair of second
class constraints.

Consider 3 primary first class constraints  (\ref{eprico}).
We choose the conformal gauge for the metric
\begin{equation}                                        \label{etigau}
    N_0=1,~~~~N_1=0,
\end{equation}
and fix the time component of the Lorentz connection
\begin{equation}                                        \label{etilor}
    B_0={\rm const}.
\end{equation}
Gauge conditions (\ref{etigau}) and (\ref{etilor}) clearly make a pairs
of second class constraints with the primary constraints. They must be
also preserved in time. The corresponding equations of motion are
\begin{eqnarray*}
	\pl_0 N_0 &=& \{N_0,H_E\}=v_0,
\\
	\pl_0 N_1 &=& \{N_1,H_E\}=v_1,
\\
	\pl_0 B_0 &=& \{B_0,H_E\}=v.
\end{eqnarray*}
So the gauge choice (\ref{etigau}) and (\ref{etilor}) fixes 3 Lagrange
multipliers in the extended Hamiltonian
$$
	v_0=v_1=v=0.
$$

Now we consider the remaining degree of freedom of the Lorentz
connection described by $Q_B$ and $P_B$. Equations of motion for
these variables are
\begin{eqnarray}                                        \label{equaqb}
    \pl_0 Q_B &=& \frac12u_0g_{11}(P_B-P_\perp)
    +\pl_1u,
\\                                                      \label{equapb}
    \pl_0 P_B &=& -(u_0+u_1)\h.
\end{eqnarray}
We choose the following gauge condition
\begin{equation}                                        \label{egacqb}
    Q_B={\rm const}
\end{equation}
It forms a pair of second class constraints with the secondary
constraints ${\cal H}_0$ and $\h$. The constraint $\h=0$
together with the equation of motion (\ref{equapb}) yields
\begin{equation}                                        \label{egacpb}
    P_B=c_B={\rm const},
\end{equation}
where $c_B$ is an arbitrary constant. As was noted in the preceeding
section addition of arbitrary constant to $P_B$ and the corresponding
constant to $P_\perp$ does not change the initial theory. So without
loss of generality we set $c_B=0$.

The gauge condition (\ref{egacqb}) must be consistent with the
equation of motion (\ref{equaqb}). This defines the Lagrange multiplier
$u$ in terms of $u_0$
\begin{equation}                                        \nonumber
    \frac12u_0g_{11}(P_B-P_\perp)+\pl_1u=0.
\end{equation}
The corresponding Hamiltonian reads
\begin{equation}                                        \label{ehamqp}
    \widehat H_E=\int d\s(u^0\widehat{\cal H}_0+u^1{\cal H}_1),
\end{equation}
where the hatted constraint $\widehat{\cal H}_0$ is obtained from the
initial constraint ${\cal H}_0$ by setting $P_B=0$
\begin{eqnarray*}
     \widehat{\cal H}_0 &=&-\frac1{2\rho}\Pi_\m\Pi^\m
     -\frac\rho2\pl_1X_\m\pl_1X^\m
\\
     &+& \frac1{2\bt}(\pl_1P)^2-\frac1{2\bt}(\pl_1P_\perp)^2
     -\pl_1PQ_\perp-\pl_1P_\perp Q
     +\pl_1^2P_\perp-g_{11}\left(\frac1{4\g}P_\perp^2+\lm\right).
\end{eqnarray*}

The constraints $\widehat{\cal H}_0$ and ${\cal H}_1$ as well as
${\cal H}_0$ and ${\cal H}_1$ form precisely the Virasoro algebra
\begin{eqnarray*}
    \{\widehat{\cal H}_0,\widehat{\cal H}'_0\} &=&
    ({\cal H}_1+{\cal H}'_1)\dl'(\s'-\s),
\\                                                      \nonumber
    \{\widehat{\cal H}_0,{\cal H}'_1\} &=&
    (\widehat{\cal H}_0+\widehat{\cal H}'_0)\dl'(\s'-\s),
\\                                                      \nonumber
    \{ {\cal H}_1,{\cal H}'_1\} &=&
    ({\cal H}_1+{\cal H}'_1)\dl'(\s'-\s).
\end{eqnarray*}

In this way we eliminate 4 unphysical degrees of freedom using 4
gauge conditions and 4 constraints. Four of the Lagrange multipliers
are also fixed by the requirement of the consistency of the gauge
conditions with the equations of motion. The net result is that we
are left with $D+2$ degrees of freedom $(X^\m,\Pi_\m)$, $(Q,P)$,
and $(Q_\perp,P_\perp)$ subjected to two first class
constraints satisfying the Virasoro algebra.
\subsection{The third canonical transformation and positive
            definiteness of the Hamiltonian}            \label{sthcan}
Although we considerably simplify the form of the constraints and
reduce their number from 6 to 2 we are still unable to provide
quantization. Therefore we will make the third canonical transformation
$$
	Q,P,Q_\perp,P_\perp\rightarrow q,p,q_\perp,p_\perp
$$
which has the peculiar feature that the quadratic part of the resulting
constraints precisely coinsides with the constraints of an ordinary
bosonic string.

Let us choose the generating function
$$
	F_3=-\pl_1Pq-\pl_1P_\perp q_\perp+\bt q_\perp\pl_1q
$$
depending on the old momenta and new coordinates. It is also nonlocal
and connects old and new variables
\begin{eqnarray}                                            \nonumber
     Q&=&-\frac{\dl F_3}{\dl P}=-\pl_1q,
\\                                                          \nonumber
     Q_\perp&=&-\frac{\dl F_3}{\dl P_\perp}=-\pl_1q_\perp,
\\                                                     \label{ethcan}
     p&=&-\frac{\dl F_3}{\dl q}=\pl_1P+\bt\pl_1q_\perp,
\\                                                          \nonumber
     p_\perp&=&-\frac{\dl F_3}{\dl q_\perp}=\pl_1P_\perp-\bt\pl_1q.
\end{eqnarray}

Then the factor $g_{11}$ in the constraints again becomes local
$$
	g_{11}=-e^{2\int Q}=-e^{-2q},
$$
the constant of integration being inessential because it can be always
normalized by choosing the appropriate value of $\s_0$ in (\ref{emetsp}).
At the same time there arises other nonlocal quantity in the constraint
$\widehat{\cal H}_0$ expressed through the new variables
\begin{equation}                                        \label{enloct}
	P_\perp=\int^\s d\s'(p_\perp'+\bt\pl_1'q')
		 =\int_{\s_2}^\s d\s'p_\perp(\s')+\bt q,
\end{equation}
with some integration constant $\s_2$.

The third canonical transformation by itself is not a one-to-one
transformation too. However, the combination of the second and the
third canonical transformation yields a one-to-one transformation for
zweibein $e_1{}^a$,$\pi^1{}_a$$\rightarrow$$q$,$q_\perp$,$p$,$p_\perp$
up to unessential addition of arbitrary constants to $q$ and
$q_\perp$.

Comparing Eqs.(\ref{ethcan}) and (\ref{esecpa}) one finds the geometric
meaning of the new variables. Coordinates $q$ and $q_\perp$ up to a
constant coinside with the conformal factor $\f$ and the angle $\om$
of $e_1{}^a$ in (\ref{ezwpar}).

In terms of the new canonical variables the constraints read
\begin{eqnarray}                                           \nonumber
     \widehat{\cal H}_0 &=&-\frac1{2\rho}\Pi_\m\Pi^\m-\frac\rho2\pl_1X_\m\pl_1X^\m
     +\frac1{2\bt}p^2+\frac\bt2(\pl_1q)^2
     -\frac1{2\bt}p_\perp^2-\frac\bt2(\pl_1q_\perp)^2
\\                                                      \label{ecothz}
     &+&\pl_1p_\perp+\bt\pl_1^2q+e^{-2q}\left(\frac1{4\g}P_\perp^2+\lm\right),
\\                                                      \label{ecotho}
     {\cal H}_1 &=& \Pi_\m\pl_1X^\m+p\pl_1q+p_\perp\pl_1q_\perp
     +\pl_1p-\bt\pl_1^2q_\perp.
\end{eqnarray}

This form of the constraints deserves some discussion.
It is quite suprising that the quadratic parts of the constraints
(\ref{ecothz}), (\ref{ecotho}) precisely coinside with the string
part of the constraints. Besides there are also linear terms and the
exponential term. The last is nonlocal because of the presence of
$P_\perp$ defined by (\ref{enloct}). It is worth mensioning that
constraints (\ref{ecothz}) and (\ref{ecotho}) look very similar to the
constraints obtained in \cite{Katana89A} where the string with dynamical
geometry was canonically formulated after fixing the conformal guage
in the equations of motion. It is not surprising because the coordinate
$q$ coinsides with the Liouville mode. I fact one can show that
constraints (\ref{ecothz}) and (\ref{ecotho}) are connected with the
constraints in Ref.~\cite{Katana89A} by canonical transformation.

Let us set for a moment $q_\perp=p_\perp=0$ naively excluding the
unphysical mode from the theory. Then the constraints read
\begin{eqnarray}                                           \nonumber
     \widehat{\cal H}_0 &=&-\frac1{2\rho}\Pi_\m\Pi^\m-\frac\rho2\pl_1X_\m\pl_1X^\m
\\                                                      \label{ecoliz}
     &+&\frac1{2\bt}p^2+\frac\bt2(\pl_1q)^2
     +\bt\pl_1^2q+\lm e^{-2q}+\frac{\bt^2}{4\g}q^2e^{-2q}
\\                                                      \label{ecolio}
     {\cal H}_1 &=& \Pi_\m\pl_1X^\m+p\pl_1q+\pl_1p.
\end{eqnarray}
This form of the constraints up to the last term in (\ref{ecoliz})
coinsides with the constraints in the string theory coupled to the
Liouville field \cite{GerNev82,BrCuTh83,Marnel83,OttWei86}. Even the
linear "improvement" term which was originally added to the Liouville
part of the constraints by hand appears here automatically. At the
same time the string with dynamical geometry is more general. The
constraints (\ref{ecothz}) and (\ref{ecotho}) which quantum algebra
will be elucidated in the second part of the paper depend from two
fields $q$ and $q_\perp$ instead of one Liouville mode. This yields
more freedom in the construction of the Fock space representation and
there does exist such representation where exponential operators
commute.

It is worth mentioning that the form of the constraints
(\ref{ecoliz}), (\ref{ecolio}) closely reminds the $\s$-model
approach to the bosonic string. Indeed, one can consider $q$ and
$q_\perp$ as two extra components of the string. The corresponding
$D+2$ dimensional string moves in a curved embedding manifold with
two timelike coordinates. The difference is that instead of $p_\perp$
one has its primitive in the exponent term.

We note that in four dimensions algebra of the constraints in general
relativity is quite similar to the Virasoro algebra and one readily
finds their stringy quadratic representation which is similar to the
constraints in the membrane model. I suppose that there also exists
the canonical transformation when the quadratic part of the
constraints will have precisely the form of the string constraints.
May be this approach will yeild consistent Fock space representation
of the constraints in quantum gravity.

New form of the constraint (\ref{ecothz}) allows one to make an
important conclusion. Namely, one of the most crucial point in
quantum field theory is positive definiteness of the canonical
Hamiltonian for physical modes. In the case of the string with
dynamical geometry it is given by $\widehat{\cal H}_0$. We readily
see that string space coordinates $X^i$, $i=1,\dots,D-1$, yield
positive contributions if and only if $\rho>0$. As noted in
Sect.\ref{scatra} the variables $q_\perp$, $p_\perp$ can be always
gauged away and are unphysical. We see that the rest of the Hamiltonian
is positive definite if $\bt,\g>0$, $\lm\ge 0$. To prove
this one has only to absorb the linear term into the quadratic one,
for example, in the momentum representation. Thus we showed that for
\begin{equation}                                        \label{epodef}
     \rho,\bt,\g>0,~~~~~~~~\lm\ge0,
\end{equation}
the canonical Hamiltonian is positive definite for all physical modes.
This statement is quite untrivial if one deals with the Hamiltonian
${\cal H}_0$ in the initial representation (\ref{econs1}).

At the end of the first part of this paper we will make the final
fourth one-to-one canonical transformation
$$
	q,q_\perp,p,p_\perp\rightarrow q^\pm,p^\pm.
$$
It is linear and defined by the generating function
$$
	F_4=-\frac1{\sqrt2}(q^++q^-)p-\frac1{\sqrt2}(q^+-q^-)p_\perp,
$$
depending on the old momenta and new coordinates. Corresponding
transformation formulas read
\begin{eqnarray*}
	q^+&=&\frac1{\sqrt2}(q+q_\perp),~~~~~~~~
	p^+=\frac1{\sqrt2}(p+p_\perp),
\\
	q^-&=&\frac1{\sqrt2}(q-q_\perp),~~~~~~~~
	p^-=\frac1{\sqrt2}(p-p_\perp).
\end{eqnarray*}
The new variables precisely coinside with the familiar light cone
variables in the string theory. One can readily write down the
constraints in terms of $q^\pm$ and $p^\pm$. They will be considered
in the next part of the paper. In this way we obtained the Hamiltonian
formulation of the string with dynamical geometry suitable for
quantization. I hope the reader is convinced that it is not so
trivial.
\section{Canonical quantization}                        \label{scanqu}
Canonical quantization means that instead of the generalized
coordinates and the conjugate momenta we must introduce noncomuting
operators acting in some representation space. As usual for quantum
field theory, we will construct the Fock space using the creation
and annihilation operators and assume the normal ordering
prescription for those operators which are nonlinear in coordinates
and momenta. Next we compute explicitly the algebra of the
quantum constraints and prove that two-dimensional gravity with
torsion by itself is anomaly free, that is there are no extra terms
as compared to the classical Poisson bracket algebra. So the
central charge in the Virasoro algebra arise only from the string
coordinates. Let us note that the ghost contribution does not taken
into account in the present paper.

To the end of this paper we consider a closed string, that is
all the fields are assumed to be periodic in space
coordinate $\s$ with the period $2\pi$. For definiteness we consider
the interval $-\pi<\s<\pi$.

The scheme of quantization is usual. We expand all the fields in
Fourier series, introduce creation and annihilation operators,
construct the Fock space, find the normal ordered expressions for the
Virasoro operators, and finally compute the quantum Virasoro algebra.
\subsection{The Fock space}                             \label{sfocks}
Here we introduce the creation and annihilation operators and
construct the Fock space for the string with dynamical geometry.
To avoid the problems arising from the space dependence of the basic
operators in the theory we define creation and annihilation
operators for the Fourier transforms of the fields.

The Fourier transform is defined as follows
\begin{eqnarray*}
	X^\m &=&\frac1{\sqrt{2\pi}}\sum_{k=-\infty}^\infty
	X_k^\m e^{\imath k\s},~~~~~~~~~~
	X_k^\m=\frac1{\sqrt{2\pi}}\int_{-\pi}^\pi d\s X^\m e^{-\imath k\s},
\\
	\Pi^\m &=&\frac1{\sqrt{2\pi}}\sum_{k=-\infty}^\infty
	\Pi_k^\m e^{\imath k\s},~~~~~~~~~~
	\Pi_k^\m=\frac1{\sqrt{2\pi}}\int_{-\pi}^\pi d\s \Pi^\m e^{-\imath k\s},
\\
	q^\pm &=&\frac1{\sqrt{2\pi}}\sum_{k=-\infty}^\infty
	q_k^\pm e^{\imath k\s},~~~~~~~~~~
	q_k^\pm=\frac1{\sqrt{2\pi}}\int_{-\pi}^\pi d\s q^\pm e^{-\imath k\s},
\\
	p^\pm &=&\frac1{\sqrt{2\pi}}\sum_{k=-\infty}^\infty
	p_k^\pm e^{\imath k\s},~~~~~~~~~~
	p_k^\pm=\frac1{\sqrt{2\pi}}\int_{-\pi}^\pi d\s p^\pm e^{-\imath k\s}.
\end{eqnarray*}
It is clear that $X_{-k}^\m=(X_k^\m)^*$ where the star denotes complex
conjugate. Similar relations hold for all other components.

The Fourier components satisfy the following nonvanishing Poisson
bracket relations
\begin{eqnarray}                                        \nonumber
	\{X_k^\m,\Pi_{\n l}\} &=& \dl_\n^\m\dl_{k,-l},
\\                                                      \label{ecpbrc}
	\{q_k^+,p_l^+\} &=& \{q_k^-,p_l^-\}=\dl_{k,-l}.
\end{eqnarray}

To construct the quantum theory we replace the components of the fields
by the operators which satisfy the canonical commutation relations
obtained from the Poisson brackets by the replacement
$\{\dots\}\rightarrow -i[\dots]$ where square brackets denote the
commutator. Then the canonical commutation relations have the form
\begin{eqnarray}                                        \nonumber
	[X_k^\m,\Pi_{\n l}] &=& \imath\dl_\n^\m\dl_{k,-l},
\\                                                      \label{eccrco}
	[q_k^+,p_l^+] &=& [q_k^-,p_l^-]=\imath\dl_{k,-l},
\end{eqnarray}
all other commutators vanishing.

Let us introduce the annihilation and creation operators for nonzero
modes $k\ne0$
\begin{eqnarray}                                        \label{eancrx}
     a_k^\m&=&\frac1{\sqrt2}\left(\sqrt{\rho|k|}X_k^\m-
     \frac i{\sqrt{\rho|k|}}\Pi_k^\m\right),
\\                                                      \label{eancx*}
     a_k^{*\m}&=&\frac1{\sqrt2}\left(\sqrt{\rho|k|}X_{-k}^\m+
     \frac i{\sqrt{\rho|k|}}\Pi_{-k}^\m\right),
\\                                                      \label{eancra}
	a_k &=& \frac1{\sqrt2}\left(\sqrt{\bt|k|}q_k^+
	+\frac\imath{\sqrt{\bt|k|}}p_k^+\right),
\\                                                      \label{eanca*}
	a_k^* &=& \frac1{\sqrt2}\left(\sqrt{\bt|k|}q_{-k}^+ -
	\frac\imath{\sqrt{\bt|k|}}p_{-k}^+\right),
\\                                                      \label{eancrb}
	b_k &=& \frac1{\sqrt2}\left(\sqrt{\bt|k|}q_k^- -
	\frac\imath{\sqrt{\bt|k|}}p_k^-\right),
\\                                                      \label{eancb*}
	b_k^* &=& \frac1{\sqrt2}\left(\sqrt{\bt|k|}q_{-k}^- +
	\frac\imath{\sqrt{\bt|k|}}p_{-k}^-\right).
\end{eqnarray}
Here and below the star denotes Hermitian conjugation of the operators.

Creation and annihilation operators for the string coordinates
(\ref{eancrx}) and (\ref{eancx*}) are introduced in the usual way.
This is done to show how the central charge arises and for comparison
with the latter construction. In section \ref{sanfre} we show how to
introduce another set of creation and annihilation operators producing
no anomaly for $X^0$ and $X^1$ components of the string.

The zero modes of the string are treated in the Schr\"odinger coordinate
representation. They act in the space of functions of $X_0^\m$. That is
the zero modes $X_0^\m$ act as the ordinary multiplication on $X_0^\m$
while the momenta act as the differention
$$
	\Pi_{\m0}=-i\frac\pl{\pl X_0^\m}.
$$
This representation of the zero modes of the string is justified by the
physical interpretation. Indeed, the zero modes $X_0^\m$ and $\Pi_{\m0}$
describe the position of the centre of mass and the total momentum of
the string. Therefore their treating in the Schr\"odinger picture is
similar to that of a point particle and describe the motion of the
string as a whole in the embedding space. At the same time the creation
operators $a_k^{*\m}$, $k\ne0$,  describe the exited states
corresponding to the oscillations of the string.

The zero modes $q_0^\pm$ and $p_0^\pm$ are treated in another way.
We introduce for them the creation and annihilation operators as well
as for nonzero modes
\begin{eqnarray}                                       \label{eancza}
	a_0 &=& \frac1{\sqrt2}\left(\sqrt\bt q_0^+ +
	\frac\imath{\sqrt\bt} p_0^+\right),
\\                                                     \label{ecreza}
	a_0^* &=& \frac1{\sqrt2}\left(\sqrt\bt q_0^+ -
	\frac\imath{\sqrt\bt} p_0^+\right),
\\                                                      \label{eanczb}
	b_0 &=& \frac1{\sqrt2}\left(\sqrt\bt q_0^- -
	\frac\imath{\sqrt\bt} p_0^-\right),
\\                                                      \label{ecrezb}
	b_0^* &=& \frac1{\sqrt2}\left(\sqrt\bt q_0^- +
	\frac\imath{\sqrt\bt} p_0^-\right).
\end{eqnarray}
These relations are obtained from those for nonzero modes by setting
$k=1$ in the right hand sides of (\ref{eancra})--(\ref{eancb*}).

This is the usual treatment of zero modes in quantum field theory.
For example, in the scalar field theory or quantum electrodynamics
the creation and annihilation operators are introduced for all modes
including the zero ones. In our case the operators $a_0^*$ and $b_0^*$
create the exited states of constant metric while the operators
$a_k^*$ and $b_k^*$, $k\ne0$, create the states with the metric
oscillating around some constant value.

At the beginning I treated the zero modes $q_0^\pm$ and $p_0^\pm$ in
the Schr\"odinger picture like they are usually treated in the quantum
Liouville theory. Both treatments yield the quantum theories free from
anomalies. Therefore the treatment of zero modes concerns only the
physical interpretation.

The formulas inverse to (\ref{eancrx})--(\ref{ecrezb}) expressing the
Fourier components through the creation and annihilation operators
look as follows
\begin{eqnarray}                                        \label{efocox}
     X_k^\m &=& \frac1{\sqrt{2\rho |k|}}(a_k^\m+a_{-k}^{*\m}),
\\                                                      \label{efocpx}
     \Pi_k^\m &=& \imath\sqrt{\frac{\rho|k|}2}(a_k^\m-a_{-k}^{*\m}),
\\                                                      \label{efocoa}
	q_k^+ &=& \frac1{\sqrt{2\bt|k|}}(a_k+a_{-k}^*),
\\                                                      \label{efocpa}
	p_k^+ &=& -\imath\sqrt{\frac{\bt|k|}2}(a_k-a_{-k}^*),
\\                                                      \label{efocob}
	q_k^- &=& \frac1{\sqrt{2\bt|k|}}(b_k+b_{-k}^*),
\\                                                      \label{efocba}
	p_k^- &=& \imath\sqrt{\frac{\bt|k|}2}(b_k-b_{-k}^*),
\end{eqnarray}
for $k\ne0$ and
\begin{eqnarray}                                        \label{efocza}
	q_0^+ &=& \frac1{\sqrt{2\bt}}(a_0+a_0^*),
\\                                                      \label{efozap}
	p_0^+ &=& -\imath\sqrt{\frac\bt2}(a_0-a_0^*),
\\                                                      \label{efoczb}
	q_0^- &=& \frac1{\sqrt{2\bt}}(b_0+b_0^*),
\\                                                      \label{efozbp}
	p_0^- &=& \imath\sqrt{\frac\bt2}(b_0-b_0^*),
\end{eqnarray}
for zero modes.

Annihilation and creation operators satisfy the canonical commutation
relations
\begin{eqnarray}                                        \label{ecranx}
     \left[a_k^\m,a_l^{*\n}\right] &=& -\et^{\m\n}\dl_{k,l},
\\                                                      \label{ecranp}
     \left[a_k,a_l^*\right] &=& \dl_{k,l},
\\                                                      \label{ecranm}
     \left[b_k,b_l^*\right] &=& -\dl_{k,l},
\end{eqnarray}
all other commutators vanishing. In formulas (\ref{ecranp}) and
(\ref{ecranm}) the indices may take zero values.

Let us note the difference in the introduction of annihilation and
creation operators. The operators $a_k$ and $a_k^*$ satisfy the same
commutation relation as the operators $a_k^i$ and $a_k^{*i}$,
$i=1,\dots,D-1$, corresponding to the space string components.
Operators $b_k$ and $b_k^*$ are similar to $a_k^0$ and $a_k^{*0}$
corresponding to the time component of the string. Note that in the
definitions (\ref{eancrx}), (\ref{eancx*}) the momenta enter
with the upper Greek index whereas the momenta conjugate to $X^\m$
(\ref{emomex}) have lower index. So for the space components
$\Pi^i=-\Pi_i$ the definition (\ref{eancrx}) has the same form as
(\ref{eancra}). This difference is obvious if one introduces
annihilation and creation operators for $p_\perp$ and $p$ because
$p_\perp$ yields negative contribution to the energy of the string and
thus must produce indefinite metric in the Fock space. Going from
$p_\perp,p$ to $p^\pm$ one can choose either the definition
(\ref{eancra}), (\ref{eancrb}) or similar definitions with
$q^+,p^+$ and $q^-,p^-$ interchanged. In fact, the only essential
point is that the commutators (\ref{ecranp}), (\ref{ecranm}) have
opposite signs.

The vacuum $|0>$ is defined as follows
$$
	a_k^\m|0>=a_k|0>=b_k|0>=0
$$
for all annihilation operators. By definition it does not depend on
$X_0^\m$
$$
	\Pi_{\m0}|0>=0.
$$
The eigenstates with total momentum $k_\m$ are defined as usual
\begin{eqnarray*}
	|0,k_\m> &=& e^{ik_\m X_0^\m}|0>,
\\
	\Pi_{\m0}|0,k_\m> &=& k_\m|0,k_\m>.
\end{eqnarray*}

The exited states of the Fock space are defined by the action of a
number of the creation operators on the vacuum with the corresponding
normalization factors
$$
	|N_1,N_2,\dots,k_\m>=\frac1{\sqrt{N_1!}}(A_1^*)^{N_1}
	\frac1{\sqrt{N_2!}}(A_2^*)^{N_2}\dots|0,k_\m>,
$$
where $A_K^*$, $K=1,2,\dots$ denotes one of the infinite set of the
creation operators $\{A_K^*\}=\{a_k^{*\m},a_k^*,b_k^*\}$. These
vectors are normalised as follows
$$
	<k'_\m,N'_1,N'_2,\dots|N_1,N_2,\dots,k_\m>=
	\pm\dl_{N_1,N'_1}\dl_{N_2,N'_2}\dots(2\pi)^D\dl^D(k_\m-k'_\m),
$$
where the integration over $X_0^\m$ is understood.

So we have constructed the Fock space. It has indefinite metric because
the action of an odd number of the creation operators $a_k^{*0}$ and
$b_k^*$ yields negative norm states. In the present paper we do
not adress the important question of elimination of these states from
the physical spectrum and concentrate our attention on the quantum
algebra of the constraints.

The constraints are quadratic in $X^\m$, $\Pi_\m$, $p^\pm$, and are
nonpolinomial in $q^\pm$ because of the presence of the exponent term.
To obtain from them the Hermitian operators acting in the Fock space
we adopt the normal ordering prescription denoted by colons. That is
all creation operators should be written on the left.
In the next sections we compute the quantum algebra of the constraints.
\subsection{The quantum constraints}                    \label{squcon}
This is a preparatory section containing the expressions of the
constraints through creation and annihilation operators and the
expressions for various contractions used in the following
calculations.

To make the computations managable let us divide the constraints into
8 pieces
\begin{eqnarray}                                        \label{expcop}
     {\cal H}^+ &=& {\cal H}_X^+ + {\cal H}_{(1)}^+ + {\cal H}_{(ab)}
     + {\cal H}_{(aa)} + {\cal H}_{(bb)} + {\cal H}_{(exp)},
\\                                                      \label{expcom}
     {\cal H}^- &=& {\cal H}_X^- + {\cal H}_{(1)}^- + {\cal H}_{(ab)}
     - {\cal H}_{(aa)} - {\cal H}_{(bb)} + {\cal H}_{(exp)}.
\end{eqnarray}
Here ${\cal H}_X^\pm$ are the familiar quadratic constraints for a string
coordinates
\begin{equation}                                        \label{econsx}
     {\cal H}_X^\pm=-:\frac14\left(
     \frac1{\sqrt\rho}\Pi_\m\mp\sqrt\rho\pl_1X_\m\right)
     \left(\frac1{\sqrt\rho}\Pi^\m\mp\sqrt\rho\pl_1X^\m\right):,
\end{equation}
where colons denote the normal ordering. The terms ${\cal H}_{(1)}^\pm$
are linear in $q^\pm$ and $p^\pm$
\begin{eqnarray}                                        \label{econsp}
     {\cal H}_{(1)}^+ &=& \frac1{\sqrt2}\pl_1p^+
     +\frac1{\sqrt2}\bt\pl_1^2q^-,
\\                                                      \label{econsm}
     {\cal H}_{(1)}^- &=& -\frac1{\sqrt2}\pl_1p^-
     +\frac1{\sqrt2}\bt\pl_1^2q^+,
\end{eqnarray}
and there is no need of normal ordering.
The term ${\cal H}_{(ab)}$ is quadratic and contains only the mixed
products of $q^+,p^+$ and $q^-,p^-$
\begin{equation}                                        \label{econab}
     {\cal H}_{(ab)}=\frac1{2\bt}:p^+p^-:
     +\frac\bt2:\pl_1q^+\pl_1q^-:.
\end{equation}
The terms ${\cal H}_{(aa)}$ and ${\cal H}_{(bb)}$ are also quadratic
but contain creation and annihilation operators of one kind
\begin{eqnarray}                                        \label{econaa}
     {\cal H}_{(aa)} &=& \frac12:p^+\pl_1q^+:,
\\                                                      \label{econbb}
     {\cal H}_{(bb)} &=& \frac12:p^-\pl_1q^-:.
\end{eqnarray}
The exponent term
\begin{equation}                                        \label{ecoexp}
     {\cal H}_{(exp)} =\frac12:e^{-\sqrt2(q^++q^-)}
     \left[\frac1{4\g}P_\perp^2-\frac1{4\g}c_B^2+\lm\right]:
\end{equation}
is defined by the corresponding Taylor series.

The terms ${\cal H}_{(ab)}$ and ${\cal H}_{(exp)}$ are common for
${\cal H}^+$ and ${\cal H}^-$, while the terms
${\cal H}_{(aa)}$ and ${\cal H}_{(bb)}$ enter with different signs.

At the classical level the constraints (\ref{expcop}) and (\ref{expcom})
satisfy the Virasoro algebra. Now the constraints become nonlinear
operators and we must compute their quantum algebra. As in the case of
ordinary bosonic string we compute the algebra of the Fourier
transformed constraints. Using the definition of creation and
annihilation operators (\ref{eancrx})--(\ref{ecrezb}) one readily finds
the explicit expressions for the coordinates and momenta
\begin{eqnarray}                                        \label{ecompx}
     X^\m &=& \frac1{\sqrt{2\pi}}X_0^\m+\frac1{\sqrt{4\pi\rho}}
	\sum_{k=1}^\infty\frac1{\sqrt k}
	\left[(a_k^\m+a_{-k}^{*\m})e^{\imath k\s}
	+(a_{-k}^\m+a_k^{*\m})e^{-\imath k\s}\right],
\\                                                      \label{ecomxp}
	\Pi^\m &=& \frac1{\sqrt{2\pi}}\Pi_0^\m
     +\imath\sqrt{\frac\rho{4\pi}}\sum_{k=1}^\infty\sqrt k
	\left[(a_k^\m-a_{-k}^{*\m})e^{\imath k\s}
	+(a_{-k}^\m-a_k^{*\m})e^{-\imath k\s}\right],
\\                                                      \label{ecompp}
	q^+ &=& \frac1{\sqrt{4\pi\bt}}(a_0+a_0^*)+\frac1{\sqrt{4\pi\bt}}
	\sum_{k=1}^\infty\frac1{\sqrt k}
	\left[(a_k+a_{-k}^*)e^{\imath k\s}
	+(a_{-k}+a_k^*)e^{-\imath k\s}\right],
\\                                                      \label{ecompa}
	p^+ &=& -\imath\sqrt{\frac\bt{4\pi}}(a_0-a_0^*)
	-\imath\sqrt{\frac\bt{4\pi}}\sum_{k=1}^\infty\sqrt k
	\left[(a_k-a_{-k}^*)e^{\imath k\s}
	+(a_{-k}-a_k^*)e^{-\imath k\s}\right],
\\                                                      \label{ecompm}
	q^- &=& \frac1{\sqrt{4\pi\bt}}(b_0+b_0^*)+\frac1{\sqrt{4\pi\bt}}
	\sum_{k=1}^\infty\frac1{\sqrt k}
	\left[(b_k+b_{-k}^*)e^{\imath k\s}
	+(b_{-k}+b_k^*)e^{-\imath k\s}\right],
\\                                                      \label{ecompb}
	p^- &=& \imath\sqrt{\frac\bt{4\pi}}(b_0-b_0^*)
	+\imath\sqrt{\frac\bt{4\pi}}\sum_{k=1}^\infty\sqrt k
	\left[(b_k-b_{-k}^*)e^{\imath k\s}
	+(b_{-k}-b_k^*)e^{-\imath k\s}\right].
\end{eqnarray}

We will also need the expression for the nonlocal field
$$
	P_\perp=\frac1{\sqrt2}\int_{\s_0}^\s d\s'(p^+ - p^-)
	-\frac\bt{\sqrt2}(q^+ +q^-)
	=\frac1{\sqrt2}(r^+ - r^-) -\frac\bt{\sqrt2}(q^+ +q^-),
$$
entering ${\cal H}_{(exp)}$. Here nonlocal fields
$r^\pm=\int_{\s_0}^\s d\s'p^\pm$ have the form
\begin{eqnarray}                                        \label{ecomrp}
	r^+ &=& -\imath\sqrt{\frac\bt{4\pi}}(a_0-a_0^*)(\s-\tilde\s)
\\                                                      \nonumber
	& &-\sqrt{\frac\bt{4\pi}}\sum_{k=1}^\infty\frac1{\sqrt k}
       \left[(a_k-a_{-k}^*)e^{\imath k\s}
	 -(a_{-k}-a_k^*)e^{-\imath k\s}\right],
\\                                                      \label{ecomrm}
	r^- &=& \imath\sqrt{\frac\bt{4\pi}}(b_0-b_0^*)(\s-\tilde\s)
\\                                                      \nonumber
	& &+\sqrt{\frac\bt{4\pi}}\sum_{k=1}^\infty\frac1{\sqrt k}
      \left[(b_k-b_{-k}^*)e^{\imath k\s}-(b_{-k}-b_k^*)e^{-\imath k\s}\right],
\end{eqnarray}
where $\tilde\s$ is the redifined integration constant. We see that
nonlocality of the canonical transformation (\ref{ecatrf}) is in fact
inessential when considering the theory of closed strings. It enters
only through the factor $(\s-\tilde\s)$ that does not yield any
difficulty in the calculations.

To perform calculations we will widely use the Wick theorem and
therefore we introduce the contraction of two linear
operators $\f$ and $\h$ as the difference between their product
and the corresponding result in normally ordered form
$$
	\cont{\f}{}{\h}=\f\h-:\f\h:.
$$
Then one easily finds the contractions for nonzero modes
\begin{eqnarray}                                        \label{econqm}
     \cont{a_k}{}{q^+}&=&\cont{q^+}{}{a_{-k}^*}=-\cont{b_k}{}{q^-}
     =-\cont{q^-}{}{b_{-k}^*}=\frac1{\sqrt{4\pi\bt k}}e^{-\imath k\s},
\\                                                      \label{econqp}
     \cont{a_{-k}}{}{q^+}&=&\cont{q^+}{}{a_k^*}=-\cont{b_{-k}}{}{q^-}
     =-\cont{q^-}{}{b_k^*}=\frac1{\sqrt{4\pi\bt k}}e^{\imath k\s},
\\                                                      \label{econrm}
     -\cont{a_k}{}{r^+}&=&\cont{r^+}{}{a_{-k}^*}=-\cont{b_k}{}{r^-}
     =\cont{r^-}{}{b_{-k}^*}=\sqrt{\frac\bt{4\pi k}}e^{-\imath k\s},
\\                                                      \label{econrp}
     \cont{a_{-k}}{}{r^+}&=&-\cont{r^+}{}{a_k^*}=\cont{b_{-k}}{}{r^-}
     =-\cont{r^-}{}{b_k^*}=\sqrt{\frac\bt{4\pi k}}e^{\imath k\s},
\end{eqnarray}
where $k>0$. We will also use the contractions of zero modes
\begin{eqnarray}                                        \label{econqz}
     \cont{a_0}{}{q^+}&=&\cont{q^+}{}{a_0^*}=-\cont{b_0}{}{q^-}
     =-\cont{q^-}{}{b_0^*}=\frac1{\sqrt{4\pi\bt}},
\\                                                      \label{econrz}
     \cont{a_0}{}{r^+}&=&-\cont{r^+}{}{a_0^*}=\cont{b_0}{}{r^-}
     =-\cont{r^-}{}{b_0^*}=\imath\sqrt{\frac\bt{4\pi}}(\s-\tilde\s).
\end{eqnarray}

The Virasoro generators are equal to the sums of the following operators
\begin{eqnarray}                                           \nonumber
     L_{Xm}^+&=&\imath\sqrt{\frac m{2\rho}}\Pi_{\m0}a_{-m}^{*\m}
     +\frac12\sum_{k=1}^{m-1}\sqrt{k(m-k)}a_{\m-k}^*a_{-m+k}^{*\m}
\\                                                      \label{evirxp}
     &-&\sum_{k=1}^\infty\sqrt{k(m+k)}a_{\m-m-k}^*a_{-k}^\m,
\\                                                          \nonumber
     L_{Xm}^-&=&-\imath\sqrt{\frac m{2\rho}}\Pi_{\m0}a_m^\m
     +\frac12\sum_{k=1}^{m-1}\sqrt{k(m-k)}a_{\m k}a_{m-k}^\m
\\                                                      \label{evirxm}
     &-&\sum_{k=1}^\infty\sqrt{k(m+k)}a_{\m k}^*a_{m+k}^\m,
\\                                                      \label{evirop}
     L_{(1)m}^+&=&m\sqrt{\frac{\pi\bt m}2}(a_m-a_{-m}^*-b_m-b_{-m}^*),
\\                                                      \label{evirom}
     L_{(1)m}^-&=&-m\sqrt{\frac{\pi\bt m}2}(a_m+a_{-m}^*-b_m+b_{-m}^*)            ],
\\                                                          \nonumber
     L_{(ab)m}&=&\frac{\sqrt m}4(a_0b_m+a_mb_0-a_0^*b_m-a_{-m}^*b_0
		     -b_0^*a_m-b_{-m}^*a_0+a_0^*b_{-m}^*+a_{-m}^*b_0^*)
\\                                                          \nonumber
     &-&\frac12\sum_{k=1}^{m-1}\sqrt{k(m-k)}
     (a_{-k}^*b_{m-k}+b_{-k}^*a_{m-k})
\\                                                      \label{evirab}
     &+&\frac12\sum_{k=1}^\infty\sqrt{k(m+k)}
     (a_{-k}b_{m+k}+a_{m+k}b_{-k}+a_k^*b_{-m-k}^*+a_{-m-k}^*b_k^*),
\\                                                          \nonumber
     L_{(aa)m}&=&\frac{\sqrt m}4
     (a_0a_m-a_0^*a_m+a_{-m}^*a_0-a_0^*a_{-m}^*)
\\                                                          \nonumber
     &+&\frac14\sum_{k=1}^{m-1}\sqrt{k(m-k)}
     (a_ka_{m-k}-a_{-k}^*a_{-m+k}^*)
\\                                                      \label{eviraa}
     &+&\frac12\sum_{k=1}^\infty\sqrt{k(m+k)}
     (a_{-m-k}^*a_{-k}-a_k^*a_{m+k}),
\\                                                          \nonumber
     L_{(bb)m}&=&-\frac{\sqrt m}4
     (b_0b_m-b_0^*b_m+b_{-m}^*b_0-b_0^*b_{-m}^*)
\\                                                          \nonumber
     &-&\frac14\sum_{k=1}^{m-1}\sqrt{k(m-k)}
     (b_kb_{m-k}-b_{-k}^*b_{-m+k}^*)
\\                                                      \label{evirbb}
     &-&\frac12\sum_{k=1}^\infty\sqrt{k(m+k)}
     (b_{-m-k}^*b_{-k}-b_k^*b_{m+k}),
\\                                                      \label{evirex}
     L_{(exp)m}&=&\int_{-\pi}^\pi d\s:{\cal H}_{exp}:e^{-\imath m\s}.
\end{eqnarray}

The last expression for $L_{(exp)m}$ cannot be written as the sum of
creation and annihilation operators in compact form. Nevertherless
calculations can be easily performed in the integral representation.

For $m=1$ finite sums in (\ref{evirxp}), (\ref{evirxm}), (\ref{evirab}),
(\ref{eviraa}), and (\ref{evirbb}) are absent.

For $m=0$ the Virasoro constraints have the form
\begin{eqnarray*}
     L_0^+&=&-\frac1{4\rho}\Pi_{\m0}\Pi_0^\m
	-\sum_{k=1}^\infty ka_{\m-k}^*a_{-k}^\m
	+\frac14(a_0b_0-a_0^*b_0-b_0^*a_0+a_0^*b_0^*)
\\
	& &+\frac12\sum_{k=1}^\infty k
	(a_{-k}b_k+a_kb_{-k}+a_k^*b_{-k}^*+a_{-k}^*b_k^*)
\\
	& &-\frac12\sum_{k=1}^\infty k
	(a_k^*a_k-a_{-k}^*a_{-k}-b_k^*b_k+b_{-k}^*b_{-k})
	+\int_{-\pi}^\pi d\s:{\cal H}_{(exp)}:,
\\
     L_0^-&=&-\frac1{4\rho}\Pi_{\m0}\Pi_0^\m
	-\sum_{k=1}^\infty ka_{\m k}^*a_k^\m
	+\frac14(a_0b_0-a_0^*b_0-b_0^*a_0+a_0^*b_0^*)
\\
	& &+\frac12\sum_{k=1}^\infty k
	(a_{-k}b_k+a_kb_{-k}+a_k^*b_{-k}^*+a_{-k}^*b_k^*)
\\
	& &+\frac12\sum_{k=1}^\infty k
	(a_k^*a_k-a_{-k}^*a_{-k}-b_k^*b_k+b_{-k}^*b_{-k})
	+\int_{-\pi}^\pi d\s:{\cal H}_{(exp)}:.
\end{eqnarray*}
Of course, arbitrary constants can be added to $L_0^\pm$ due to an
ambiquity in the operator ordering. We drop this possibility at present
because we will not discuss in detail the physical interpretation of
the Fock space in the present paper. The constraint $L_0$ then reads
\begin{eqnarray}                                        \nonumber
     L_0&=&L_0^+ + L_0^-=-\frac1{2\rho}\Pi_{\m0}\Pi_0^\m
	-\sum_{k=1}^\infty k(a_{\m k}^*a_k^\m+a_{\m-k}^*a_{-k}^\m)
\\                                                      \nonumber
	&+&\frac12(a_0b_0-a_0^*b_0-b_0^*a_0+a_0^*b_0^*)
\\                                                      \label{etoten}
	&+&\sum_{k=1}^\infty k
	(a_{-k}b_k+a_kb_{-k}+a_k^*b_{-k}^*+a_{-k}^*b_k^*)
	+2\int_{-\pi}^\pi d\s:{\cal H}_{(exp)}:.
\end{eqnarray}
Its quadratic part corresponding to $q^\pm$ is undiagonal and yields
the problem of interpretation of the states. We briefly discuss this
point at the end of Section \ref{sanfre}. Here we note that the
constructed Fock space contains two unphysical modes besides the
physical ones and $L_0$ is not the energy of the string but just the
constraint. Therefore one has no reason for the diagonality of $L_0$.

Expressions for the operators with negative $m<0$ are easily obtained
from those of positive $m>0$ by Hermitian conjugation
$$
	L_{-m}=(L_m)^*.
$$
They are not written down explicitly for the sake of space.

So we got expressions for all terms entering the Virasoro
constraints. The terms $L_{(1)m}^\pm$ are linear, while $L_{Xm}^\pm$,
$L_{(ab)m}$, $L_{(aa)m}$, $L_{(bb)m}$ are belinear in creation and
annihilation operators. The exponent term $L_{(exp)m}$ is
nonpolinomial and is understood as the normally ordered power series.
\subsection{General remark}                             \label{sgenre}
Here we make the general remark concerning calculations of commutators
of quantum operators. Suppose we have two polynomial operators
$:p_n^Nq_m^M:$ and $:p_k^Kq_l^L:$ written as the normally ordered product
on canonical coordinates $q_m$, $q_l$ and conjugate momenta $p_n$,
$p_k$. Here $N$, $M$, $K$, and $L$ denote the corresponding powers.

Classical expressions of these operators have the following Poisson
bracket
$$
	\{p_n^Nq_m^M,p_k^Kq_l^L\}=MKp_n^Nq_m^{M-1}p_k^{K-1}q_l^L\dl_{m,-k}
	-NLp_n^{N-1}q_m^Mp_k^Kq_l^{L-1}\dl_{n,-l},
$$
where for definitness we used the basic Poisson brackets (\ref{ecpbrc}).

Let us consider the commutator
\begin{equation}                                        \label{ecommu}
	\left[:p_n^Nq_m^M:,:p_k^Kq_l^L:\right].
\end{equation}
Due to the Wick theorem the right hand side of this commutator contains
normal products with zero number of contraction, single contractions,
double contractions, and so on. If there were no contractions then the
operators would commute. If there were only single contractions then
one would obtain exactly the classic result multiplied by $\imath$.
Indeed, consider all terms in the right hand side of (\ref{ecommu})
containing single contractions
\begin{eqnarray}                                        \nonumber
     \left[:p_n^Nq_m^M:,:p_k^Kq_l^L:\right]&\stackrel{(1)}{=}&
     NK:p_n^{N-1}q_m^Mp_k^{K-1}q_l^L:
     \left(\cont{p_n}{}{p_k}-\cont{p_k}{}{p_n}\right)
\\                                                      \nonumber
     & &+NL:p_n^{N-1}q_m^Mp_k^Kq_l^{L-1}:
     \left(\cont{p_n}{}{q_l}-\cont{q_l}{}{p_n}\right)
\\                                                      \nonumber
     & &+MK:p_n^Nq_m^{M-1}p_k^{K-1}q_l^L:
     \left(\cont{q_m}{}{p_k}-\cont{p_k}{}{q_m}\right)
\\                                                      \label{ecomon}
     & &+ML:p_n^Nq_m^{M-1}p_k^Kq_l^{L-1}:
     \left(\cont{q_m}{}{q_l}-\cont{q_l}{}{q_m}\right),
\end{eqnarray}
where superscript $(1)$ denotes that only single contractions are
used in the Wick theorem and where elementary contractions are given by
\begin{eqnarray*}
	\cont{p_n}{}{p_k}&=&\frac{\bt|n|}2\dl_{n,-k},
\\
	\cont{p_n}{}{q_l}&=&-\frac{\imath}2\dl_{n,-l},
\\
	\cont{q_l}{}{p_n}&=&~\frac{\imath}2\dl_{n,-l},
\\
	\cont{q_m}{}{q_l}&=&\frac1{2\bt|m|}\dl_{m,-l},
\end{eqnarray*}
For definitness but without loss of generality we consider here the
contractions for $q_k^+$ and $p_k^+$ (\ref{efocoa}), (\ref{efocpa}).
Then the first and the last terms
in (\ref{ecomon}) dissappear and one is left with the classic result
$$
	\left[:p_n^Nq_m^M:,:p_k^Kq_l^L:\right]\stackrel{(1)}{=}
	\imath MK:p_n^Nq_m^{M-1}p_k^{K-1}q_l^L:\dl_{m,-k}
	-\imath NL:p_n^{N-1}q_m^Mp_k^Kq_l^{L-1}:\dl_{n,-l}.
$$

It is clear that no extra terms as compared to the classical result
appear if only single contractions are taken into account. Thus single
contractions exactly reproduce the classical Poisson bracket relations
up to the factor $\imath$. This result is readily generalised to
arbitrary polinomials in $q,p$.

The terms containing two or more contractions yield the quantum
corrections. If at the classical level some set of nonlinear operators
obeyed the closed Poisson bracket algebra, then at the quantum level
this algebra could be broken by the quantum corrections producing the
anomaly. A set of first class constraints reflects the gauge symmetry
of the action and the existence of anomaly means that the symmetry is
broken at the quantum level. The famous example of such a theory is
the ordinary bosonic string where the conformal symmetry is broken
by the central charge of the Virasoro algebra arising from double
contractions in the Wick theorem. In critical dimension $D=26$ the
central charge can be compensated in the enlarged Fock space by
ghost modes.

To compute the anomaly we must consider only the terms containing two
or more contractions. In the following sections we show that in spite
of the nonpolinomial constraints in the theory of the string with
dynamical geometry the only anomaly is the anomaly of ordinary bosonic
string. The computation will be devided in two parts. First, we
will consider the diagonal terms and next the
crossterms in the commutators.
\subsection{Diagonal terms}                             \label{sdiagt}
Let us consider the commutators of $L_{Xm}^\pm$, $L_{(ab)m}$,
$L_{(aa)m}$, $L_{(bb)m}$, and $L_{(exp)m}$ with themselves. The
operators $L_{(1)m}^\pm$ are linear and due to the general remark their
commutators  produce no anomalous terms. Therefore they are not
considered in what follows.

Consider the commutator
$$
	[L_{Xm}^+,L_{Xn}^+].
$$
The terms containing one creation and one annihilation operators do
not produce any anomaly because one of the two contractions is always
zero. So double contractions can arise only within the terms included
in the finite sum (\ref{evirxp}). For $m,n\ge0$ there are no double
contractions and therefore the anomaly is absent. The commutator
$$
	[L_{Xm}^+,L_{X-n}^+],
$$
where $m,n>0$ does contain double contractions arising from the finite
sums
\begin{eqnarray}                                        \nonumber
	[L_{Xm}^+,L_{X-n}^+]&\stackrel{(2)}{=}&\frac14
	\left[\sum_{k=1}^{m-1}\sqrt{k(m-k)}a_{\m-k}^*a_{-m+k}^{*\m},
	\sum_{l=1}^{n-1}\sqrt{l(n-l)}a_{\n-l}a_{-n+l}^{\n}\right]
\\                                                      \nonumber
	&\stackrel{(2)}{=}&-\frac14\sum_{k=1}^{m-1}\sum_{l=1}^{n-1}
	\sqrt{k(m-k)l(n-l)}\left[
	\cont{a_{\n-l}}{}{a_{\m-k}^*}\cont{a_{-n+l}^\n}{}{a_{-m+k}^{*\m}}
	+\cont{a_{\n-l}}{}{a_{-m+k}^{*\m}}\cont{a_{-n+l}^\n}{}{a_{\m-k}^*}
	\right]
\\                                                       \label{ecench}
	&=&-\frac D2\sum_{k=1}^{m-1}k(m-k)\dl_{m,n}
	=-\frac D{12}m(m^2-1)\dl_{m,n}.
\end{eqnarray}
This is the famous expression for the central charge in the Virasoro
algebra. The zero modes do not change its value irrespective of their
normal ordering prescription because $L_{Xm}^+$ does not contain
$X_0^\m$.

The same calculations yield the central charge in the commutator
$$
	[L_{Xm}^-,L_{X-n}^-]\stackrel{(2)}{=}
	\frac D{12}m(m^2-1)\dl_{m,n}.
$$
It differs from (\ref{ecench}) only by the sign because nontrivial
contractions arise when $L_{Xm}^-$ stands on the left.

In contrast to the $L_{Xm}^\pm$ the finite sum in $L_{(ab)m}$ does
not yield any contribution to the anomaly. The suspicious terms are
only those containg zero modes and entering the infinite sum.
For $m,n\ge0$, the anomalous terms in the commutator
$$
	\left[L_{(ab)m},L_{(ab)n}\right]
$$
are clearly absent. Nontrivial calculations occur only in the
commutator
\begin{eqnarray*}
	& &\left[L_{(ab)m},L_{(ab)-n}\right]
	\stackrel{(2)}{=}\frac{\sqrt{mn}}{16}\left[
	\cont{a_0}{}{a_0^*}\cont{b_m}{}{b_n^*}+\cont{b_0}{}{b_0^*}
	\cont{a_m}{}{a_n^*}
	-\cont{a_0}{}{a_0^*}\cont{b_{-n}}{}{b_{-m}^*}
	-\cont{b_0}{}{b_0^*}\cont{a_{-n}}{}{a_{-m}^*}\right]
\\
	&&+\frac14\sum_{k,l=1}^\infty\sqrt{k(m+k)l(n+l)}
\\
      &&\times\left[\cont{a_{-k}}{}{a_{-l}^*}\cont{b_{m+k}}{}{b_{n+l}^*}
	+\cont{a_{m+k}}{}{a_{n+l}^*}\cont{b_{-k}}{}{b_{-l}^*}
	-\cont{a_l}{}{a_k^*}\cont{b_{-n-l}}{}{b_{-m-k}^*}
	-\cont{a_{-n-l}}{}{a_{-m-k}^*}\cont{b_l}{}{b_k^*}\right]
\\
	&&=\frac m{16}(-2\dl_{m,n}+2\dl_{m,n})
	+\frac14\sum_{k=1}^\infty k(m+k)(-2\dl_{m,n}+2\dl_{m,n})=0.
\end{eqnarray*}
So all anomalous terms cancell. This happened because the operator
$L_{(ab)m}$ contains terms quadratic in creation operators along
with the terms quadratic in annihilation operators.

The terms yielding nontrivial double contractions in $L_{(aa)m}$ are
those containing the zero mode and entering the finite sum. Nonzero
double contractions arise only in the commutator
\begin{eqnarray*}
	& &\left[L_{(aa)m},L_{(aa)-n}\right]
	\stackrel{(2)}{=}\frac{\sqrt{mn}}{16}\left[
	\cont{a_0}{}{a_0^*}\cont{a_m}{}{a_n^*}
	-\cont{a_0}{}{a_0^*}\cont{a_{-n}}{}{a_{-m}^*}\right]
\\
	&&+\frac14\sum_{k=1}^{m-1}\sum_{l=1}^{n-1}\sqrt{k(m-k)l(n-l)}\left[
	\cont{a_k}{}{a_l^*}\cont{a_{m-k}}{}{a_{n-l}^*}
	-\cont{a_{-l}}{}{a_{-k}^*}\cont{a_{-n+l}}{}{a_{-m+k}^*}\right]
\\
	&&=\frac m{16}(\dl_{m,n}-\dl_{m,n})
	+\frac14\sum_{k=1}^{m-1}k(m-k)(\dl_{m,n}-\dl_{m,n})=0,
\end{eqnarray*}
where $m,n>0$. Again all anomalous terms cancell.

Similar calculations show that all anomalous terms in the diagonal
commutator
$$
	\left[L_{(bb)m},L_{(bb)n}\right]
$$
cancell for all values of $m$ and $n$. We omit here the details
for the sake of space.

The anomalous terms do not appear in the commutator
$$
	\left[L_{(exp)m},L_{(exp)n}\right]
$$
for another reason. Here one encounts infinite number of contractions
but all off them are zero. Indeed, operator $L_{(exp)m}$ depends only
from two combinations of the fields $q^++q^-$ and $p^+-p^-$. Easy
analysis show that
\begin{equation}                                        \label{econex}
    \cont{(q^++q^-)}{}{(q^++q^-)}=\cont{(q^++q^-)}{}{(p^+-p^-)}
    =\cont{(p^+-p^-)}{}{(p^+-p^-)}=0.
\end{equation}
Thus the exponential terms clearly commute due to the Wick theorem. The
essential point here is that the creation operators for $q^+$ and
$q^-$ produce the states of opposite norm. Only in this case
contractions (\ref{econex}) equal zero.

So we see the difference between the quantum Liouville theory and
quantum two-dimensional gravity with torsion. Instead of
one field the latter theory contains a pair of fields which contribute
to the energy with opposite signs. These fields can be arranged in
such a way that the exponents commute. In this way
one of the main trouble of quantum Liouville theory is avoided.
\subsection{Crossterms}                                 \label{scross}
To compute the crossterms commutators one should perform more tedious
calculations as compared to the diagonal commutators. It is clear
that $L_{Xm}^+$ commute with $L_{Xm}^-$ and both of them commute with
the operators $L_{(ab)m}$, $L_{(aa)m}$, $L_{(bb)m}$, and $L_{(exp)m}$.
The crossterms commutators
$$
	\left[L_{(ab)m},L_{(aa)n}\right]\stackrel{(2)}{=}
	\left[L_{(ab)m},L_{(bb)n}\right]\stackrel{(2)}{=}
	\left[L_{(aa)m},L_{(bb)n}\right]\stackrel{(2)}{=}0
$$
also yield no anomaly because there are unsufficient number of
creation and annihilation operators of the same kind. So we have to
consider the crossterms between $L_{(ab)m}$, $L_{(aa)m}$, $L_{(bb)m}$
and the exponent term $L_{(exp)m}$. Here we must consider only double
contractions because one of the operators is necessarily quadratic.

Let us consider the crossterms
\begin{equation}                                        \label{eabexp}
     \left[L_{(ab)m},L_{(exp)n}\right]
     +\left[L_{(exp)m},L_{(ab)n}\right]=
     \left[L_{(ab)m},L_{(exp)n}\right]-(m\leftrightarrow n)
\end{equation}
which enter the commutator $\left[L_m^+,L_n^+\right]$ precisely in this
combination.
To show how the cancellation occurs we consider first the commutators
of the terms containing the zero modes in $L_{(ab)m}$ and the
cosmological term in $L_{(exp)n}$
$$
     \left[\frac{\sqrt m}4(a_0b_m+a_mb_0+a_0^*b_{-m}^*+a_{-m}^*b_0^*),
     \frac\lm2\int_{-\pi}^\pi d\s:e^{-\sqrt2(q^++q^-)}:e^{-\imath n\s}\right]
     -(m\leftrightarrow n).
$$
Expanding the operator exponent in a Taylor series, taking all possible
double contractions and then gathering the exponent again one obtains
\begin{eqnarray*}
	& &\frac{\lm\sqrt m}4\int_{-\pi}^\pi d\s:e^{-\sqrt2(q^++q^-)}:
	e^{-\imath n\s}
\\
      &&\times\left[\cont{a_0}{}{q^+}\cont{b_m}{}{q^-}
	+\cont{a_m}{}{q^+}\cont{b_0}{}{q^-}
	-\cont{q^+}{}{a_0^*}\cont{q^-}{}{b_{-m}^*}
	-\cont{q^+}{}{a_{-m}^*}\cont{q^-}{}{b_0^*}\right]-(m\leftrightarrow n)
\\
	& &=\frac\lm4\int_{-\pi}^\pi d\s:e^{-\sqrt2(q^++q^-)}:
	e^{-\imath n\s}\left[-\frac1{4\pi\bt}e^{-\imath m\s}
	+\frac1{4\pi\bt}e^{-\imath m\s} \right]-(m\leftrightarrow n)=0,
\end{eqnarray*}
where contractions (\ref{econqm})--(\ref{econrz}) have been used.
So the cancellation occurs inside the square brackets in the last
expression before the antisymmetrization. Computing the commutator of
quadratic terms containing zero
modes in $L_{(ab)m}$ with whole exponential term one must also take
into account the double contractions with $(r^+-r^-)$ and double
mixed contractions: the first with $(q^++q^-)$ and the second with
$(r^+-r^-)$. Double contractions with $(q^++q^-)$ when one factor
$(q^++q^-)$ is taken from the exponent and the other is taken from
$P_\perp^2$ term cancell like in the previous case. Double
contractions with $(r^+-r^-)$ cancell precisely in the same way as those
with $(q^++q^-)$. Double mixed contractions cancell in another way.
That is, computing the commutator (\ref{eabexp}) one obtains the term
\begin{eqnarray*}
	&&\sqrt m\int_{-\pi}^\pi d\s
	:e^{-\sqrt2(q^+-q^-)}(P_\perp-c_B):e^{-\imath n\s}
\\
      &&\times\left[
	-\cont{a_0}{}{q^+}\cont{b_m}{}{r^-}
	+\cont{a_0}{}{r^+}\cont{b_m}{}{q^-}
	-\cont{a_m}{}{q^+}\cont{b_0}{}{r^-}
	+\cont{a_m}{}{r^+}\cont{b_0}{}{q^-}\right.
\\
	& &\left.+\cont{q^+}{}{a_0^*}\cont{r^-}{}{b_{-m}^*}
	-\cont{r^+}{}{a_0^*}\cont{q^-}{}{b_{-m}^*}
	+\cont{q^+}{}{a_{-m}^*}\cont{r^-}{}{b_0^*}
	-\cont{r^+}{}{a_{-m}^*}\cont{q^-}{}{b_0^*}\right]
	-(m\leftrightarrow n)
\\
	&&=\int_{-\pi}^\pi d\s:e^{-\sqrt2(q^++q^-)}(P_\perp-c_B):
	\left[\frac1\pi-\frac\imath\pi(\s-\tilde\s)\right]e^{-\imath(m+n)\s}
	-(m\leftrightarrow n)=0,
\end{eqnarray*}
where we have dropped insignificant constant factor. It is cancelled
after antisymmetrization in $m,n$. So we have proved
that double contractions of the first term in $L_{(ab)m}$ with
$L_{(exp)n}$ produce no anomaly. Now we are left to consider double
contractions of the infinite sum in (\ref{evirab}) because finite sum
can yield only single nontrivial contractions. Using contractions
(\ref{econqm})--(\ref{econrp}) one easily shows that all double
contractions with $(q^++q^-)^2$, $(q^++q^-)(r^+-r^-)$, and
$(r^+-r^-)^2$ cancell. We omit the corresponding calculations. Thus
for $m,n>0$ the crossterm commutator equals zero
$$
     \left[L_{(ab)m},L_{(exp)n}\right]-(m\leftrightarrow n)=0.
$$
This is the exact result because single contractions also yield zero
as follows from Poisson brackets calculation. One can check by similar
calculations that this result holds for all values of $m$ and $n$.

Let us consider the next crossterms commutator
\begin{equation}                                        \label{eaaexp}
     \left[L_{(aa)m},L_{(exp)n}\right]-(m\leftrightarrow n)
\end{equation}
for $m,n>0$. The first term in $L_{(aa)m}$ containing the zero mode
produce the following double contractions with $(q^++q^-)^2$,
$(q^++q^-)(r^+-r^-)$, and $(r^+-r^-)^2$
\begin{eqnarray*}
	& &\sqrt m\left[\cont{a_0}{}{q^+}\cont{a_m}{}{q^+}
	+\cont{q^+}{}{a_0^*}\cont{q^+}{}{a_{-m}^*}\right]e^{-\imath n\s}
	-(m\leftrightarrow n)=\frac1{4\pi\bt}e^{-\imath(m+n)\s}
	-(m\leftrightarrow n)=0,
\\
	& &\sqrt m\left[\cont{a_0}{}{q^+}\cont{a_m}{}{r^+}
	+\cont{a_0}{}{r^+}\cont{a_m}{}{q^+}
	+\cont{q^+}{}{a_0^*}\cont{r^+}{}{a_{-m}^*}
	+\cont{r^+}{}{a_0^*}\cont{q^+}{}{a_{-m}^*}
	\right]e^{-\imath n\s}
	-(m\leftrightarrow n)
\\
	& &=\left[-\frac1{4\pi}+\imath\frac{\s-\tilde\s}{4\pi}
	+\frac1{4\pi}
	-\imath\frac{\s-\tilde\s}{4\pi}\right]e^{-\imath(m+n)\s}
	-(m\leftrightarrow n)=0,
\\
	& &\sqrt m\left[\cont{a_0}{}{r^+}\cont{a_m}{}{r^+}
	+\cont{r^+}{}{a_0^*}\cont{r^+}{}{a_{-m}^*}\right]e^{-\imath n\s}
	-(m\leftrightarrow n)
\\
      &&=-\imath\frac{\bt(\s-\tilde\s)}{2\pi}
	e^{-\imath(m+n)\s}-(m\leftrightarrow n)=0,
\end{eqnarray*}
and all of them cancell. Here we omit the integration over $\s$ because
cancellation occures under the integral sign and drop constant factors.
Next we have to consider double contractions between the finite sum in
$L_{(aa)m}$ and $L_{(exp)n}$. Here we must be carefull because the
upper limit of summation depends on $m$. Straightforward calculations
show that two terms arising from double contractions with $(q^++q^-)^2$,
$(q^++q^-)(r^+-r^-)$, and $(r^+-r^-)^2$ survive
\begin{eqnarray}                                          \nonumber
	& &\sum_{k=1}^{m-1}\sqrt{k(m-k)}\left[
	\cont{a_k}{}{q^+}\cont{a_{m-k}}{}{q^+}
	+\cont{q^+}{}{a_{-k}^*}\cont{q^+}{}{a_{-m+k}^*}\right]e^{-\imath n\s}
	-(m\leftrightarrow n)
\\                                                      \label{ecraqq}
      &&=\sum_{k=1}^{m-1}
	\frac1{2\pi\bt}e^{-\imath(m+n)\s}-(m\leftrightarrow n)\ne0,
\\                                                        \nonumber
	& &\sum_{k=1}^{m-1}\sqrt{k(m-k)}\left[
	\cont{a_k}{}{q^+}\cont{a_{m-k}}{}{r^+}
	+\cont{q^+}{}{a_{-k}^*}\cont{r^+}{}{a_{-m+k}^*}\right]e^{-\imath n\s}
	-(m\leftrightarrow n)
\\                                                      \label{ecraqr}
      &&=\sum_{k=1}^{m-1}
	\left(-\frac1{4\pi}+\frac1{4\pi}\right)e^{-\imath(m+n)\s}
	-(m\leftrightarrow n)=0,
\\                                                         \nonumber
	& &\sum_{k=1}^{m-1}\sqrt{k(m-k)}\left[
	\cont{a_k}{}{r^+}\cont{a_{m-k}}{}{r^+}
	+\cont{r^+}{}{a_{-k}^*}\cont{r^+}{}{a_{-m+k}^*}\right]e^{-\imath n\s}
	-(m\leftrightarrow n)
\\                                                      \label{ecrarr}
      &&=\sum_{k=1}^{m-1}
	\frac{\bt}{2\pi}e^{-\imath(m+n)\s}-(m\leftrightarrow n)\ne0.
\end{eqnarray}
Here for brevity we also do not write the integral over $\s$.
The first and the third terms do not cancell because of the upper limit
in the sum. So we have considered all double contractions in the
crossterms commutator (\ref{eaaexp}) and they lead to a nonzero result.

Let us consider the last crossterms commutator
\begin{equation}                                        \label{ebbexp}
     \left[L_{(bb)m},L_{(exp)n}\right]-(m\leftrightarrow n)
\end{equation}
for $m,n>0$. Calculations in this case are very similar to the
previous ones. We obtain precisely the same results with one very
important difference. The finite sum in $L_{(bb)m}$ has opposite
sign to that of $L_{(aa)m}$ and one obtains the terms which exactly
cancell (\ref{ecraqq}) and (\ref{ecrarr}). So we have proved that for
$m,n>0$ the crossterms commutator
$$
     \left[L_{(aa)m}+L_{(bb)m},L_{(exp)n}\right]-(m\leftrightarrow n)=0
$$
equals zero. This results can be generalised for all values of $m$ and
$n$ by direct calculations which are omitted here.

Thus we have proved that all double contractions in crossterms
commutators cancell and there is no quantum corrections to the
classical algebra of the constraints in pure two-dimensional gravity
with torsion. The only anomaly comes from the string coordinates, and
the quantum algebra of the constraints coincides with that of
ordinary bosonic string
\begin{eqnarray*}
     \left[L_m^+,L_n^+\right]&=&-(m-n)L_{m+n}^+
     -\frac D{12}m(m^2-1)\dl_{m,-n},
\\
     \left[L_m^+,L_n^-\right]&=&0,
\\
     \left[L_m^-,L_n^-\right]&=&~(m-n)L_{m+n}^-
     +\frac D{12}m(m^2-1)\dl_{m,-n},
\end{eqnarray*}

So the central charge in the Virasoro algebra of the string with
dynamical geometry precisely coinside with that of an ordinary
bosonic string if we use the standart representation for the
string coordinates. In the following section we construct new
Fock representation for a bosonic string where the value of the
central charge differs by two.
\subsection{Anomaly free string in two dimensions}      \label{sanfre}
The canonical transformation converting $q,p$ and $q_\perp,p_\perp$
into $q^\pm,p^\pm$ before the introduction of creation and annihilation
operators is a crucial point for the construction of the anomaly free
representation of the Virasoro algebra for pure two-dimensional gravity
with torsion. If we constructed the Fock space directly from $q,p$ and
$q_\perp,p_\perp$ then we would obtain the anomalous theory.
There would arise not only the central charge but the algebra would be
completely broken. For example, the exponential terms would produce
other exponential terms with different exponents. As a by-product of
our construction we proved that in the Fock space for $q^\pm,p^\pm$
the central charge produced by quadratic terms in (\ref{ecothz}) and
(\ref{ecotho}) which have precisely the same form as the constraints
for string coordinates does not appear at the quantum level. This
sugests new representation for an ordinary bosonic string.

Let us make the canonical transformation to the light cone variables
$$
	X^0,X^1,\Pi_0,\Pi_1\rightarrow X^\pm,\Pi^\pm,
$$
with the generating function
$$
	F_5=-\frac1{\sqrt2}(X^+-X^-)\Pi_0-\frac1{\sqrt2}(X^++X^-)\Pi_1
$$
depending on new coordinates and old momenta. Other string
coordinates remain unchanged. New coordinates and momenta are related
to the old ones by the formulas
\begin{eqnarray*}
	X^+&=&\frac1{\sqrt2}(X^1+X^0),~~~~~~~~
	\Pi^+=\frac1{\sqrt2}(\Pi_1+\Pi_0),
\\
	X^-&=&\frac1{\sqrt2}(X^1-X^0),~~~~~~~~
	\Pi^-=\frac1{\sqrt2}(\Pi_1-\Pi_0).
\end{eqnarray*}

We introduce annihilation and creation operators $c,d$ and
$c^*,d^*$ for the Fourier transformed light cone variables
\begin{eqnarray}                                        \label{elcann}
     X_k^+&=&\frac1{\sqrt{2\rho |k|}}(d_k+d_{-k}^*),~~~~~~
     X_k^-=\frac1{\sqrt{2\rho |k|}}(c_k+c_{-k}^*),
\\                                                      \label{elccre}
     \Pi_k^+&=&-\imath\sqrt{\frac{\rho |k|}2}(d_k-d_{-k}^*),~~~~~~
     \Pi_k^-=\imath\sqrt{\frac{\rho |k|}2}(c_k-c_{-k}^*),
\end{eqnarray}
where $k\ne0$. They satisfy canonical commutation relations
\begin{equation}                                        \label{ecacor}
     -\left[c_k,c_l^*\right]=\left[d_k,d_l^*\right]=\dl_{k,l},
\end{equation}
and are related to the old creation and annihilation operators
(\ref{eancrx}), (\ref{eancx*}) by linear transformation
\begin{eqnarray}                                        \label{eneodc}
     c_k&=&\frac1{\sqrt2}(a_{-k}^{*1}-a_k^0),~~~~~~~~
     d_k=\frac1{\sqrt2}(a_k^1+a_{-k}^{*0}),
\\                                                      \label{eneola}
     c_k^*&=&\frac1{\sqrt2}(a_{-k}^1-a_k^{*0}),~~~~~~~~
     d_k^*=\frac1{\sqrt2}(a_k^{*1}+a_{-k}^0).
\end{eqnarray}
Here important point is that the canonical commutation relations of
$c_k$ and $c_k^*$ have negative sign as well as that of $a_k^0$ and
$a_k^{*0}$.

In the present section for brevity we omit the zero modes because they
do not contribute to the central charge.

In terms of new creation and annihilation operators the $X^0$, $X^1$
part of the constraints for $m>0$ up to zero modes read
\begin{equation}                                        \label{enewxc}
     L_{Xm}^\pm=L_{(cd)m}\pm(L_{(cc)m}+L_{(dd)m}),
\end{equation}
where
\begin{eqnarray}                                           \nonumber
     L_{(cd)m}&=&-\frac12\sum_{k=1}^{m-1}\sqrt{k(m-k)}
     (d_{-k}^*c_{m-k}+c_{-k}^*d_{m-k})
\\                                                      \label{evircd}
     &+&\frac12\sum_{k=1}^\infty\sqrt{k(m+k)}
     (d_{-k}c_{m+k}+d_{m+k}c_{-k}+d_k^*c_{-m-k}^*+d_{-m-k}^*c_k^*),
\\                                                          \nonumber
     L_{(cc)m}&=&-\frac14\sum_{k=1}^{m-1}\sqrt{k(m-k)}
     (c_kc_{m-k}-c_{-k}^*c_{-m+k}^*)
\\                                                      \label{evircc}
     &-&\frac12\sum_{k=1}^\infty\sqrt{k(m+k)}
     (c_{-m-k}^*c_{-k}-c_k^*c_{m+k}),
\\                                                          \nonumber
     L_{(dd)m}&=&\frac14\sum_{k=1}^{m-1}\sqrt{k(m-k)}
     (d_kd_{m-k}-d_{-k}^*d_{-m+k}^*)
\\                                                      \label{evirdd}
     &+&\frac12\sum_{k=1}^\infty\sqrt{k(m+k)}
     (d_{-m-k}^*d_{-k}-d_k^*d_{m+k}).
\end{eqnarray}
This form of the constraints have precisely the same form as
$L_{(ab)}\pm(L_{(aa)}+L_{(bb)})$ after replacement $a,b\rightarrow d,c$.
Therefore formal commutators have no anomalous term as compared to the
classical Poisson brackets.

This is an intriguing observation, but we must pay for the absence of
the central charge. In fact, the constraints (\ref{evircd})--(\ref{evirdd})
are not well defined on the basis of the new Fock space because they
produce the states with infinite norm. So they are unbounded operators
and one must be carefull to give a precise meaning of the commutators.
At present the calculations must be considered only as the formal
manipulations with the operator symbols. We hope that there exists more
rigorous foundation of the present calculations.

In contrast to the ordinary representation one can impose the whole set
of quantum constraints on the physical subspace
$$
	L_m^\pm|phys>=0
$$
for all $m$. At the formal level the physical space is not empty.
For example, the state
$$
	|ph>=\exp\left[\pm\sum_{l=1}^\infty
	(c_l^*c_{-l}^*+d_l^*d_{-l}^*\right]|0>
$$
is annihilated by all $L_{Xm}^\pm$
$$
     L_{(cd)m}|ph>=L_{(cc)m}|ph>=L_{(dd)m}|ph>=0.
$$
It has infinite norm and its interpretation is obscure. But the mere
fact of its existence seems to be interesting.

One may try also to find new Fock representation for the remaining
space components of the string by introducing complex string
coordinates like $X^2\pm\imath X^3$ and corresponding creation and
annihilation operators. Unfortunately, this approach does not solve
the problem of anomaly. This shows that different signs in the
canonical commutation relations (\ref{ecacor}) are of prime importance.

In the new representation the quantum Virasoro algebra has the form
\begin{eqnarray*}
     \left[L_m^+,L_n^+\right]&=&-(m-n)L_{m+n}^+
     -\frac{D-2}{12}m(m^2-1)\dl_{m,-n},
\\
     \left[L_m^+,L_n^-\right]&=&0,
\\
     \left[L_m^-,L_n^-\right]&=&~(m-n)L_{m+n}^-
     +\frac{D-2}{12}m(m^2-1)\dl_{m,-n}.
\end{eqnarray*}
We see that in $D=2$ the string with dynamical geometry is anomaly
free. That is the central charge in the Virasoro algebra equals zero.

Let us briefly comment this model. If one considers the ordinary
bosonic string in two dimensions then one will find that the theory
describes no continious degree of freedom because two Virasoro
constraints and two gauge conditions eliminate both degrees of freedom
$X^0$ and $X^1$. The situation in pure two-dimensional gravity with
torsion is the same: one can always eliminate all continious degrees
of freedom from the theory. At the same time the united theory of
two-dimensional string with dynamical geometry describes two degrees
of freedom and is anomaly free. The physical degrees of freedom
in this case are the space component of the string $X^1$ and the
conformal factor of the metric $g_{\al\bt}$. They yield positive
definite Hamiltonian if the coupling constants in the Lagrangian
satisfy (\ref{epodef}).

At first sight the problem of tachyon will be
successfully solved for two dimensional string with dynamical geometry.
Indeed, in the ordinary approach the tachyon arises due to the
constant $\al_0$ in the quantum expression for the constraint $L_0$.
The negative value of $\al_0$ is intirely connected to the anomaly in
the Virasoro algebra, for example, in the BRST approach. In our case
the anomaly is absent and I see no reason for arising of the tachyon.
This question as well as the spectrum and proper interpretation of the
new Fock representation will be considered in detail elswhere.
\section{Conclusion}                                    \label{sconcl}
Let us make a few remarks concerning construction of quantum gravity.
General relativity as well as other theories with general coordinate
invariance contains a set of nonlinear constraits. If one tries to
quantize the theory in a perturbative approach then one finds that
the algebra of the constraints do not closes in each order of the
perturbation theory already at the classical level. This raises a
question of consistency of the perturbative approach because to
maintain general coordinate invariance at the quantum level one must
retain all orders up to the order of the constraints. In four
dimensional general relativity the constraints are nonpolinomial and
therefore one must retain infinite number of perturbations to maintain
the invariance at the quantum level. Therefore a nonperturbative
approach to quantum gravity is desirable.

In the present paper we quantized two-dimensional gravity with torsion
without any approximation. It is quite suprising that the algebra of
nonpolinomial constraints do have a Fock space representation where
the quantum algebra of the constraints coinsides with the classical
Poisson bracket algebra. That is the theory is anomaly free neglecting
the ghost contribution. To obtain a true anomaly free gauge theory one
has to construct anomaly free representation for ghost fields. This
representation appears to be different from the usual one and we hope
that it exists.

To construct the representation of the constraints we have widely used
the canonical transformations corresponding to extracting the
longitidinal and transversal parts from the zweibein. This is a
nontrivial problem in a curved background and the canonical
transformation is nonlocal. The remarkable point is that the quadratic
parts of the resulting constraints precisely coinside with the
constraints of ordinary bosonic string. This interesting observation
may be a common feature of gravity theories. Hopefully, a similar
canonical transformation separating physical and unphysical modes
exists in arbitrary dimensional space-time, and this is the proper
way of approaching the problem of quantum gravity.

The approach used in the present paper demonstrates power and
flexibility of the canonical transformations in separation of physical
and unphysical modes in the theory. At the same time we have
shown that even linear transformation of creation and annihilation
operators can result in unitary inequivalent Fock space
representations of the nonlinear operators.
Nonlocality of the canonical transformation is not new and disastrous
feature of the theory. For example, to seperate explicitly longitudinal
part of the electromagnetic potential one also has to perform
nonlocal transformation. There arises a question whether there exists
and how to quantize the string with dynamical geometry in terms of the
initial variables in the Lagrangian. I hope a detailed elucidation
of this problem in this  relatively simple case will shed a new light
on quantum canonical transformations and on the whole problem of
quantum gravity.

The author would like to express his gratitude to I.~V.~Volovich
for numerous enlightening discussions and to S.~L.~Lyakhovich,
D.~J.~Schwarz and Th.~Strobl for the conversations on Hamiltonian
formulation of two-dimensional gravity with torsion and quantization.


\begin{thebibliography}{10}

\bibitem{GrScWi87}
M.~B. Green, J.~H. Schwarz, and E.~Witten.
\newblock {\em Superstring theory}, volume 1,2.
\newblock Cambridge U.P., Cambridge, 1987.

\bibitem{BriHen88}
L.~Brink and M.~Henneaux.
\newblock {\em Principles of String Theory}.
\newblock Plenum Press, New York and London, 1988.

\bibitem{KatVol86}
M.~O. Katanaev and I.~V. Volovich.
\newblock String model with dynamical geometry and torsion.
\newblock {\em Phys.\ Lett.}, 175B(4):413--416, 1986.

\bibitem{KatVol90}
M.~O. Katanaev and I.~V. Volovich.
\newblock Two-dimensional gravity with dynamical torsion and strings.
\newblock {\em Ann.\ Phys.}, 197(1):1--32, 1990.

\bibitem{KatVol92}
M.~O. Katanaev and I.~V. Volovich.
\newblock Theory of defects in solids and three-dimensional gravity.
\newblock {\em Ann.\ Phys.}, 216(1):1--28, 1992.

\bibitem{Katana90}
M.~O. Katanaev.
\newblock Complete integrability of two-dimensional gravity with dynamical
  torsion.
\newblock {\em J.\ Math.\ Phys.}, 31(4):882--891, 1990.

\bibitem{KumSch92A}
W.~Kummer and D.~J. Schwarz.
\newblock General analytic solution of {$R^2$}--gravity with dynamical torsion
  in two dimensions.
\newblock {\em Phys.\ Rev.\ D}, 45(8):3628--3635, 1992.

\bibitem{Solodu93}
S.~N. Solodukhin.
\newblock Two-dimensional black holes with torsion.
\newblock {\em Preprint JINR E2--93--33}, page 10 pp., 1993.

\bibitem{MiGrObTrHe93}
E.~W. Mielke, F.~Gronwald, Yu.~N. Obukhov, R.~Tresguerres, and F.~W. Hehl.
\newblock Towards complete integrability of two dimensional {P}oincar\'e gauge
  gravity.
\newblock {\em Preprint Cologne-thp-1993-H6}, page 35 pp., 1993.

\bibitem{AkKiRi88}
K.~J. Akdeniz, A.~Kizilers\"{u}, and E.~Rizao\v{g}lu.
\newblock Instanton and eigenmodes in a two-dimensional theory of gravity with
  torsion.
\newblock {\em Phys.\ Lett.}, B215(1):81--83, 1988.

\bibitem{Katana91}
M.~O. Katanaev.
\newblock Conformal invariance, extremals, and geodesics in two-dimensional
  gravity with torsion.
\newblock {\em J.\ Math.\ Phys.}, 32(9):2483--2496, 1991.

\bibitem{Katana93A}
M.~O. Katanaev.
\newblock All universal coverings of two-dimensional gravity with torsion.
\newblock {\em J.\ Math.\ Phys.}, 34(2):700--736, 1993.

\bibitem{GrKuPrSc92}
H.~Grosse, W.~Kummer, P.~Pre\v{s}najder, and D.~J. Schwarz.
\newblock Novel symmetry of non-{E}insteinian gravity in two dimensions.
\newblock {\em J.\ Math.\ Phys.}, 33:3892--3900, 1992.

\bibitem{Strobl93}
T.~Strobl.
\newblock All symmetries of non-{E}insteinian gravity in $d=2$.
\newblock {\em Int.\ J.\ Mod.\ Phys.}, A8:1383, 1993.

\bibitem{Katana89A}
M.~O. Katanaev.
\newblock String with dynamical geometry. {H}amiltonian analysis in conformal
  gauge.
\newblock {\em Theor.\ Math.\ Phys.}, 80(2):239--252, 1989.
\newblock [In Russian].

\bibitem{AkDaKi92}
K.~J. Akdeniz, \"{O}.~F. Dayi, and A.~Kizilers\"{u}.
\newblock {\em Mod.\ Phys.\ Lett.}, A7:1757, 1992.

\bibitem{ArDeMi62}
R.~Arnowitt, S.~Deser, and S.~W. Misner.
\newblock The dynamics of general general relativity.
\newblock In L.~Witten, editor, {\em Gravitation: an introduction to current
  research}, New York -- London, 1962. John Wiley \& Sons, Inc.

\bibitem{SchStr92}
P.~Schaller and T.~Strobl.
\newblock Canonical quantization of non-{E}insteinian gravity and the problem
  of time.
\newblock {\em Preprint TUW--92--13}, page 19 pp., 1992.

\bibitem{KumSch92B}
W.~Kummer and D.~J. Schwarz.
\newblock Renormalization of {$R^2$}--gravity with dynamical torsion in $d=2$.
\newblock {\em Nucl.\ Phys.\ B}, 382:171--186, 1992.

\bibitem{HaiKum92}
F.~Haider and W.~Kummer.
\newblock Quantum functional integration of non-{E}insteinian gravity in $d=2$.
\newblock {\em Preprint TUW--92--15}, page 16 pp., 1992.

\bibitem{IkeIza93A}
N.~Ikeda and K.-J. Izawa.
\newblock Quantum gravity with dynamical torsion in two dimensions.
\newblock {\em Prog.\ Theor.\ Phys.}, 89:223, 1993.

\bibitem{IkeIza93B}
N.~Ikeda and K.-J. Izawa.
\newblock Gauge theory based on quadratic lie algebras and {2D} gravity with
  dynamical torsion.
\newblock {\em Preprint RIMS--911}, page 12pp., 1993.

\bibitem{HeTeZa90}
M.~Henneaux, C.~Teitelboim, and J.~Zanelli.
\newblock Gauge invariance and degree of freedom count.
\newblock {\em Nucl.\ Phys. B}, 332(1):169--188, 1990.

\bibitem{GerNev82}
J.-L. Gervais and A.~Neveu.
\newblock Dual string spectrum in {P}olaykov's quantization. {(II). M}ode
  separation.
\newblock {\em Nucl.\ Phys.}, B209(1):125--145, 1982.

\bibitem{BrCuTh83}
E.~Braaten, Th. Curtright, and C.~Thorn.
\newblock An exact operator solution of the quantum {L}iouville field theory.
\newblock {\em Ann.\ Phys.}, 147:365--416, 1983.

\bibitem{Marnel83}
R.~Marnelius.
\newblock Canonical quantization of {P}olyakov's string in arbitrary
  dimensions.
\newblock {\em Nucl.\ Phys.}, B211(1):14--28, 1983.

\bibitem{OttWei86}
H.~J. Otto and G.~Weigt.
\newblock Construction of exponential {L}iouville field operators for closed
  string models.
\newblock {\em Z.\ Phys.\ C}, 31(2):219--228, 1986.

\end{thebibliography}
\end{document}